\documentclass[twocolumn,showpacs,preprintnumbers,amsmath,amssymb]{revtex4}
\usepackage{amsmath,amsfonts,latexsym,amssymb,graphics,epsfig,subfigure,color,makeidx}
\usepackage{xcolor}

\begin{document}

\title{Charged spinning black holes as accelerators of spinning particles}

\author{Yu-Peng Zhang\footnote{zhangyupeng14@lzu.edu.cn},
        Bao-Min Gu\footnote{gubm15@lzu.edu.cn},
        Shao-Wen Wei \footnote{ weishw@lzu.edu.cn},
        Jie Yang \footnote{ yangjiev@lzu.edu.cn},
        Yu-Xiao Liu \footnote{liuyx@lzu.edu.cn, corresponding author}
}
\affiliation{Institute of Theoretical Physics, Lanzhou University, Lanzhou 730000, China}

\begin{abstract}

It is well known that some black holes can act as accelerators for particles without spin. Recently, there are some works considering collision of two spinning particles in the background of Schwarzschild and Kerr black holes and it was shown that {the center-of-mass energy of the test particles is related to the spin}. In this paper we extend the results to some more general cases. We consider Kerr-Newman black holes as accelerators for spinning particles. We derive the center-of-mass energy of the spinning particles and use numerical method to investigate how the center-of-mass energy is affected by the properties of the black holes and spinning particles.

\end{abstract}


\maketitle

\section{Introduction}\label{scheme1}

Ba$\tilde{n}$ados, Silk, and West first showed that extremal Kerr black holes can act as accelerators of particles in 2009, and the center-of-mass (CM) energies of two test particles can be arbitrary high if the collision occurs near the horizon~\cite{Banados2009}.
However, the spin of a black hole should be less than $0.998M$ ($M$ is the mass of the black hole) if the limitations from astrophysics are considered~\cite{Thorne1974}, and the authors in Refs. \cite{Berti2009,Jacobson2010} showed that Planck-scale collisions of two spinless particles cannot occur near the horizon of a Kerr black hole with spin less than $0.998M$. We know that when a black hole has charge besides spin the motions of test particles will change, so the CM energy of two test particles will depend on both the spin and charge of the black hole. In Ref. \cite{Wei2010} the authors obtained the CM energy for the collision of two test particles in the background of a charged spinning black hole and found that it could diverge with some conditions. Some other black holes as accelerators of spinless particles have also been studied systematically (see Refs. \cite{Wei2010a,Li2011,Harada2011,Zhu2011,Sadeghi2012,Frolov2012,Patil2012,Abdujabbarov2013,Tursunov2013,Galajinsky2013,Harada2014,Zaslavskii2014,Sultana2015,Sultana2015a,Ghosh2015} for examples).

It is well known that the motion of a test particle with nonzero spin will not be the geodesic anymore~\cite{Hanson1974,phdthesis,Hojman1977,Zalaquett2014}. Recently, Armaza, Ba$\tilde{n}$ados, and Koch \cite{Armaza2016} investigated the collision of two test particles with nonzero spin near the horizon of a Schwarzschild black hole and found that the CM energy of the two particles can be divergent out of the horizon under some conditions. Multiple scattering for this kind of collision was taken into account in Ref.~\cite{Zaslavskii1603.09353}. The work \cite{Armaza2016} was extended to the Kerr black hole case in Ref.~\cite{Guo2016}, which showed that the CM energy near the horizon can be arbitrarily high for an extremal Kerr black hole, but can not for a non-extremal one. However, it is more interesting to investigate the case out of the horizon of a general black hole. In this paper we will study the collision of two spinning test particles in the equatorial plane of a Reissner-Nordstrom (RN) black hole, Kerr black hole, and Kerr-Newman (KN) black hole, respectively. Some new results will be found. For example, the CM energy for two spinning test particles out of the horizon of a non-extremal Kerr black hole might also be divergent, and the area of the  divergent region in the spin-angular space of the particle increases with the spin of the black hole. { In Refs. \cite{phdthesis,Armaza2016,Hojman2013,Deriglazov2015,Deriglazov2015a} show that the velocity vector $u^\mu$ and the canonical momentum vector $P^\mu$ of the spinning test particle are not parallel, and the velocity vector $u^\mu$ might transform to be spacelike from timelike along the trajectory. So it is important to clarify the relation between the divergent region and the superluminal region in spin-angular space.}

Our paper is organized as follows. In Sec. \ref{review} we review the equations of motion for a spinning test particle in curved spacetime. In Sec. \ref{scheme1} we obtain the four momentum of a spinning test particle based on Ref. \cite{Hojman1977} and calculate the CM energy for two spinning test particles in the KN background. In Sec. \ref{scheme2} we consider the possibility that RN black holes act as spinning test particle accelerators and investigate the characters of the CM energy as a function of the particle spin $s$ and orbital angular momentum $l$ with different values of the black hole charge by using a numerical method. In Sec. \ref{scheme3} we consider the case of Kerr black holes and investigate the characters of the CM energy as a function of the spin $s$ and total angular orbital mentum $l$ with different values of the black hole spin $a$. In Sec. \ref{scheme4} we consider the case of extremal KN black holes. {In Sec. \ref{scheme5} we investigate the velocity of the spinning test particles and find that the spinning test particles actually can not reach the divergent radius out of the horizon because of the superluminal motion of the spinning test particles. Finally, we give a brief summary and conclusion in Sec. \ref{Conclusion}.}

\section{Review of equation of motion for a spinning particle}{\label{review}}

In this section, we review the Lagrangian mechanism to solve the equations of motion for a spinning test particle based on Ref. \cite{phdthesis}. We use $x^\mu$ to represent the position for the spinning test particle (or a top), and the orientation of the top is defined by the orthonormal tetrad $e_{(\alpha)}^\mu$, which satisfies $g^{\mu\nu}=e^{\mu}_{(\alpha)}e^{\nu}_{(\beta)}\eta^{(\alpha\beta)}$ and $\eta^{(\alpha\beta)}=\text{diag}(+1, -1, -1, -1)$. The four velocity is defined by
    \begin{equation}
    u^{\mu}\equiv\frac{dx^\mu}{d\lambda}.\label{velocity}
    \end{equation}
The angular velocity tensor $\sigma^{\mu\nu}$ is defined by
    \begin{equation}
    \sigma^{\mu\nu}\equiv\eta^{(\alpha\beta)}e_{(\alpha)}^\mu\frac{D e_{(\beta)}^\nu}{D\lambda}=-\sigma^{\nu\mu},\label{angularvelocity}
    \end{equation}
where $\frac{D e_{(\beta)}^\nu}{D \lambda}$ is the covariant derivative and has the form
    \begin{equation}
    \frac{D e_{(\beta)}^\nu}{D \lambda}\equiv\frac{de_{(\beta)}^\nu}{d\lambda}+\Gamma^\nu_{\rho\tau}e_{(\beta)}^\rho u^\tau.\label{covatiantderivative}
    \end{equation}

A possible Lagrangian should be constructed in terms of invariant quantities. As stated in Refs. \cite{phdthesis,Hojman1977}, there are four independent invariants:
\begin{eqnarray}
a_1&=&u^\mu u_\mu,\nonumber\\
a_2&=&\sigma^{\mu\nu}\sigma_{\mu\nu}=-\texttt{tr}(\sigma^2),\nonumber\\ a_3&=&u_\alpha\sigma^{\alpha\beta}\sigma_{\beta\gamma}u^\gamma,\nonumber\\ a_4&=&g_{\mu\nu}g_{\rho\tau}g_{\alpha\beta}g_{\gamma\delta}\sigma^{\delta\mu}\sigma^{\nu\rho}
\sigma^{\tau\alpha}\sigma^{\beta\gamma}.
\end{eqnarray}
Then the Lagrangian can be expressed as $\mathcal{L}=\mathcal{L}(a_1,a_2,a_3,a_4)$. With the supplementary condition $S^{\mu\nu}P_\nu=0$ \cite{phdthesis} the final equations of motion for the spin top are
    \begin{equation}
    \frac{D P^{\mu}}{D \lambda}=-\frac{1}{2}R^\mu_{\nu\alpha\beta}u^\nu S^{\alpha\beta},\label{equationmotion1}
    \end{equation}
and
    \begin{equation}
    \frac{D S^{\mu\nu}}{D \lambda}=S^{\mu\lambda}\sigma_\lambda^\nu-\sigma^{\mu\lambda}S^\nu_\lambda=P^\mu u^\nu-u^\mu P^\nu,\label{equationmotion2}
    \end{equation}
where the conjugate momentum vector $P_\mu$ and spin tensor $S_{\mu\nu}$ are defined by
    \begin{equation}
    P_\mu\equiv\frac{\partial \mathcal{L}}{\partial u^\mu},~~~~~~S_{\mu\nu}\equiv\frac{\partial \mathcal{L}}{\partial \sigma^{\mu\nu}}=-S_{\nu\mu}.
    \end{equation}
We can see that the spinning particle does not follow the geodesics.

\section{Spinning particles in Charged Spinning Black Hole Background}\label{scheme1}

In this section we consider the spinning test particles moving in the KN black hole background, which is described by the Boyer-Lindquist coordinates
    \begin{eqnarray}
    ds^2&=&-\frac{\Delta}{\rho^2}(dt-a~\sin^2\theta d\phi)^2+\frac{\rho^2}{\Delta}dr^2+\rho^2 d\theta^2\nonumber\\
    &&+\frac{\sin^2\theta}{\rho^2}[(r^2+a^2)d\phi-adt]^2,\label{metric}
    \end{eqnarray}
where $\Delta=r^2-2Mr+a^2+Q^2$ and $\rho^2=r^2+a^2\cos^2\theta$, with $Q$ being the charge per unit rest mass of the black hole and $a$ the angular momentum per unit rest mass. The outer and inner  horizons of the KN black hole are
    \begin{equation}
    r_{\pm}=1\pm\sqrt{1-(a^2+Q^2)},\label{fermionaction}
    \end{equation}
where we have chosen $M=1$ for simplicity.
This requires a constraint on $a$ and $Q$:
    \begin{equation}
    a^2+Q^2\leq 1,\label{fermionaction}
    \end{equation}
where $``="$ corresponds to the extremal black hole with one degenerate horizon.

For simplicity, we only consider the orbits in the equatorial plane with $\theta=\frac{\pi}{2}$, for which the non-vanishing  components of the conjugate momentum are \cite{Hojman1977}
    \begin{eqnarray}
    P^t&=&\frac{m^3}{\Theta \Xi}\Bigg\{r^2 a\bar{j}(2Mr-Q^2)-\bar{e} r^6+2a~\bar{e}-\bar{j})Q^2 r^2\bar{s}\nonumber\\
    &&-r^2a^2\bar{e}(2Mr+r^2-Q^2)+a^2(a~\bar{e}-\bar{j})(Q^2-M~r)\bar{s}\nonumber\\
    &&+(\bar{j}-3a~\bar{e}) M r^3 \bar{s}\Bigg\},\label{memantumpt}
    \end{eqnarray}
    \begin{eqnarray}
    P^\phi&=&\frac{m^3}{\Theta \Xi}\Bigg(a~\bar{e}~r^2 (Q^2-2Mr)+a^2 \bar{e}(Q^2-Mr)\bar{s}+r^2Q^2\nonumber\\
    &&+a\bar{j} (Mr-Q^2)\bar{s}+r^2 (r^2-2Mr)(\bar{e}\bar{s}-\bar{j})\Bigg),\label{memantumpp}
    \end{eqnarray}
    and
    \begin{eqnarray}
    (P^r)^2&=&\frac{m^6}{r^2 \Xi^2}\Bigg[r^6 \bigg(2 M  r^3 -r^2\left(\bar{j}^2+a^2-a^2\bar{e}^2+ Q^2\right)\nonumber\\
    &&+j_e^2 (2M~ r-Q^2)+(\bar{e}^2-1)r^4\bigg)\nonumber\\
    &&+2 r^4\bar{s}\left(-aQ_rj_e^2+2\bar{e}j_eQ^2 r^2-3\bar{e}j_eMr^3+\bar{e}\bar{j} r^4\right)\nonumber\\
    &&-Q_r^2\Theta \bar{s}^4+r^2 \bar{s}^2a^2 Q_r\bar{e}^2 (Q_r+2 r^2)\nonumber\\
    &&+r^2 \bar{s}^2\bigg(\bar{j}^2 Q_r^2-2 a \bar{e}\bar{j}Q_r\left(Q_r+r^2\right)-a^2 Q_r2 r^2\nonumber\\
    &&-r^2 \left(Q^2+r^2-2Mr\right) \left(\bar{e}^2 r^2+2Q_r\right)\bigg)\Bigg].\label{memantumpr}
    \end{eqnarray}
Here $\bar{e}=\frac{e}{m}$, $\bar{s}=\frac{s}{m}$, and {$\bar{j}=\frac{j}{m}=\frac{l}{m}+\frac{s}{m}$} are the energy, spin angular momentum, and total angular momentum per unit mass of the test particle, and
 \begin{eqnarray}
    \Theta~ &=& a^2+Q^2-2Mr+r^2,\\
    \Xi~ &=& m^2 r^4+(Q^2-Mr)m^2\bar{s}^2,\\
    Q_r &=& Q^2-Mr,\quad j_e=\bar{j}-a\bar{e}.
    \end{eqnarray}
The corresponding CM energy has the following form:
    \begin{equation}
    E_{\text{cm}}^2=-(P_1+P_2)^2=m_1^2+m_2^2-2P_1\cdot P_2.\label{centerofmassenergy}
    \end{equation}
By inserting Eqs. (\ref{memantumpt}--\ref{memantumpr}) into (\ref{centerofmassenergy}) and considering the two  particles having the same mass $m$, we obtain the CM energy of the two spinning test particles
    \begin{eqnarray}
    E_{\text{cm}}^2=\frac{2m^2}{\Delta^2\Xi_1\Xi_2}K,\label{centerofmassenergyjiexijie}
    \end{eqnarray}
where $\Xi_i=m^2 r^4+(Q^2-Mr)m^2\bar{s}_i^2$ ($i=1, 2$) and the expression of $K$ is very long and we do not list it here. It is naive to see that when $r\to r_{+}$ the value of $\Delta$ is zero and the CM energy might diverge, but the  numerator of Eq. (\ref{centerofmassenergyjiexijie}) near the horizon also vanishes. The divergence of the CM energy near the horizon for the extremal KN black hole depends on the spin angular momenta  and total orbital angular momenta of the two particles. We can see that the case $\Xi_i=0$ is more interesting because the CM energy is likely to diverge outside the horizon of the black hole.

It can be seen that our result for the case with $s_i=0$ is the same as the result of the KN black hole as an accelerator for spinless particles in Ref. \cite{Wei2010}. For the case with $Q=0$ and $\bar{s}_i=0$, our result can also reduce to the case of the Kerr black hole as an accelerator for the spinless particles in Ref. \cite{Banados2009}. The case with $a=Q=0$ describes the Schwarzschild black hole as { an accelerator} of spinning test particles given in Ref. \cite{Armaza2016}.

\section{Reissner-Nordstrom black hole case}\label{scheme2}

Firstly, we let $a=0$ and consider the RN black hole as a spinning test particles accelerator. The corresponding CM energy is
    \begin{eqnarray}
    (E^{\text{RN}}_{\text{cm}})^2=\frac{2 K_{1}}{\left(Q^2+(r-2) r\right)\Xi_1\Xi_2},\label{centerofmassenergyrn}
    \end{eqnarray}
where the expression of $K_1$ is given in appendix A. Because the RN { black hole} does not have spin angular momentum, we choose $\bar{j}_1=-\bar{j}_2=\bar{j}>0$ and $\bar{s}_1=-\bar{s}_2=\bar{s}>0$, and then the CM energy becomes
    \begin{equation}
    (E^{\text{RN}}_{\text{cm}})^2=\frac{4 m^2 \left\{r^6(\bar{j}-\bar{s})^2+\left[r^4+\bar{s}^2(Q^2-r)\right]^2\right\}}{\left[r^4+\bar{s}^2 (Q^2-r)\right]^2}.\label{cmenergyrn}
    \end{equation}
For the extremal RN black hole, the CM energy near the horizon is
    \begin{equation}
    E^{\text{RN}}_{\text{cm}}|_{r\to r_+}=\frac{m}{2}\sqrt{16+\big[(\bar{j}_1-\bar{s}_1)-(\bar{j}_2-\bar{s}_2) \big]^2}.\label{extremalrnhorizon}
    \end{equation}
It is similar to the case of spinless particles in Schwarzschild black hole, for which $E^{\text{Schw}}_{\text{cm}}|_{r\to 2}=\frac{m}{2}\sqrt{16+\big[(\frac{l_1}{m})-(\frac{l_2}{m}) \big]^2}$ \cite{Banados2009}. We can see that the CM energy near the horizon of an extremal RN black hole  with arbitrary spin $\bar{s}_i$ and total angular momentum $\bar{j}_i$ is always finite.

\begin{figure}[!htb]
    \includegraphics[width=0.23\textwidth]{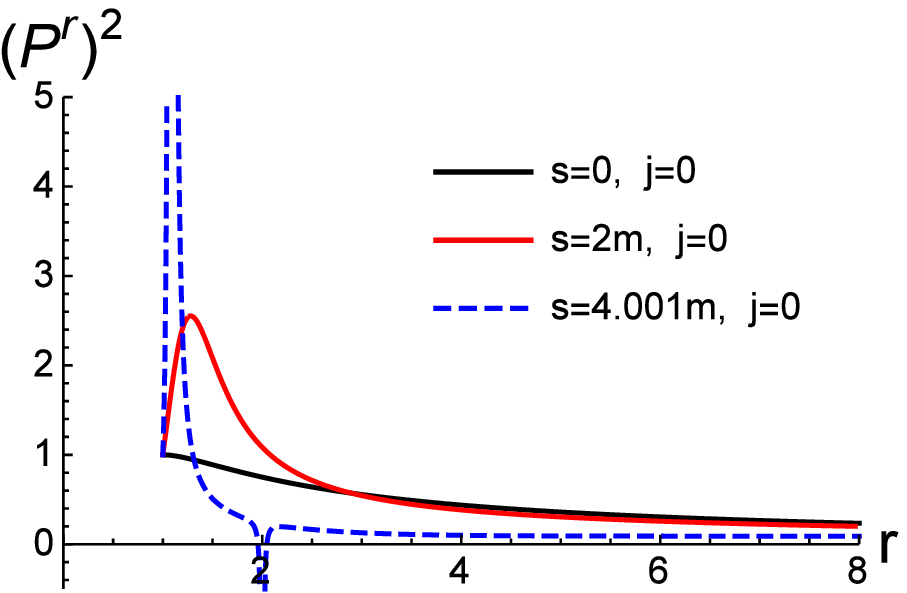}
    \includegraphics[width=0.23\textwidth]{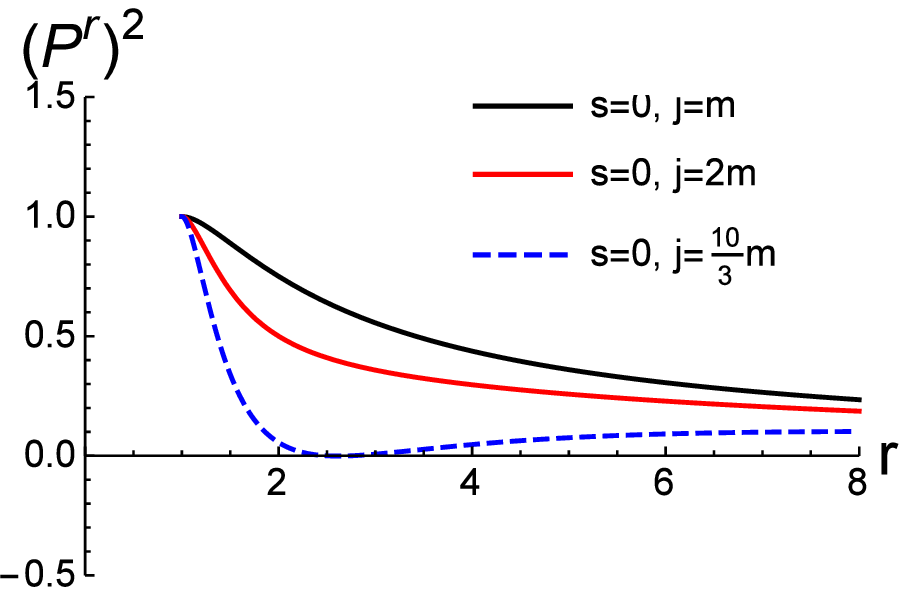}
    \vskip -4mm \caption{Plots of the $(P^r)^2$ with different $s$ and $j$ for the extremal RN black hole ($Q=1$).}
    \label{turnpointextremalrn}
    \end{figure}

For the non-extremal RN black hole the CM energy might diverge when $m^2 r^4+m^2\bar{s}^2(Q^2-r)=0$ and $\bar{j}\neq \bar{s}$. The only thing we should worry about is that the particles are inside the horizon when the CM energy reaches infinity, which means the CM energy actually can never be infinite. There are two cases that the CM energy could be infinite:
\begin{itemize}
  \item  $0<Q\leq\frac{\sqrt{15}}{4}$ and $\bar{s}\geq\frac{16 Q^3}{3 \sqrt{3}}$;
  \item  $\frac{\sqrt{15}}{4}<Q<1$ and $\bar{s}>\sqrt{\frac{\left(\sqrt{1-Q^2}+1\right)^3 }
             {1-Q^2}}$.
\end{itemize}
However, it is easy to know that if the spin $s$ and the total angular momentum $j$ are too large, the particles can not reach the horizon. That is to say, there is a turning radius where the particles
turn around and go back. Because the velocity $u^r$ is propositional to the momentum $P^r$, we use the zero point of $P^r$ to define the turning point. At the turning radius, we have $(P^r)^2= 0$. If this turning radius is larger than the divergent radius, then the CM energy will not diverge. For the extremal RN black hole, the plots of $(P^r)^2$ with respect to $s$ and $j$ are shown in Fig. \ref{turnpointextremalrn}. We see that the particle can reach the horizon and the CM energy could be divergent near the horizon.

\begin{figure}[!htb]
    \includegraphics[ width=0.23\textwidth]{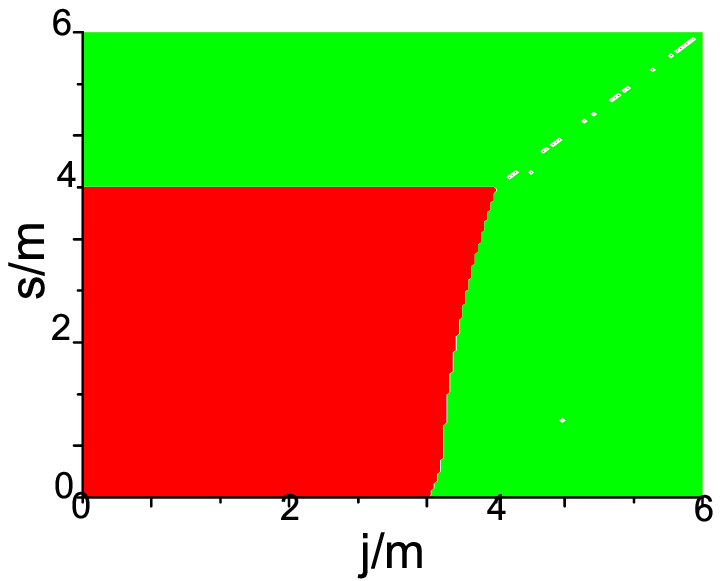}
    \includegraphics[ width=0.23\textwidth]{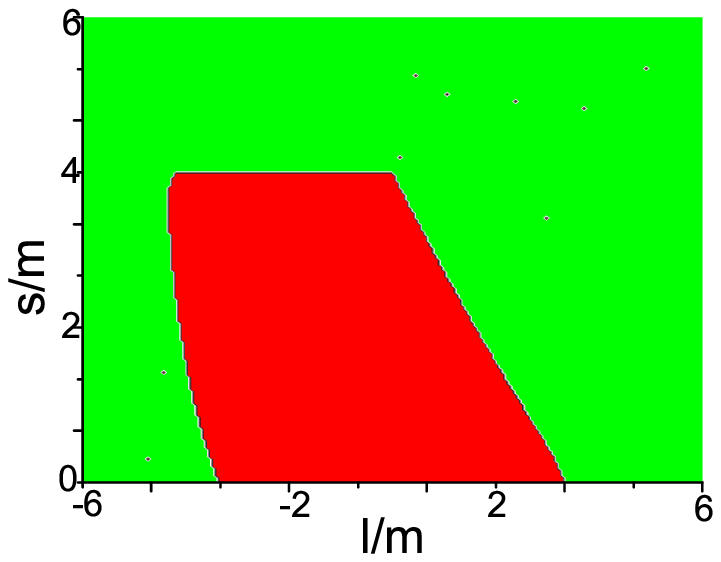}
    \vskip -4mm \caption{Plots of the characters of the CM energy for the extremal RN black hole as a function of the total angular momentum $j$ (left figure) or orbital angular momentum $l$ (right figure) and spin $s$. In red region (lower left region in the left figure and lower region in the right figure) the particles can reach the horizon while in green region (upper right region in the left figure and upper region in the right figure) they can not.  The step length of spin, orbital angular momentum, and total angular momentum are $\frac{6}{400}$, $\frac{12}{400}$, and $\frac{6}{400}$, respectively. The particles move from $r=80$ to the horizon of the black hole and the step length is $\frac{80}{12000}$.}
    \label{extremalrnsj}
    \end{figure}

The CM energy depends on three parameters, $s$, $j$, and $Q$. We numerically analyze the turning point radius $r_t$ and the divergent radius $r_d$ to determine the parameter space. For the extremal RN black hole, the numerical results are shown in Fig. \ref{extremalrnsj}. We can see that for the extremal RN black hole, the spinning test particles can reach the horizon if the spin is about in the range $-4m<s<4m$ ($l=0$) and the orbital angular momentum satisfies $-\frac{10m}{3}<l<\frac{10m}{3}$ ($s=0$), which is the same as the result in Fig. \ref{turnpointextremalrn}.

\begin{figure}[!htb]
    \subfigure[~$Q=0$]{
    \includegraphics[ width=0.23\textwidth]{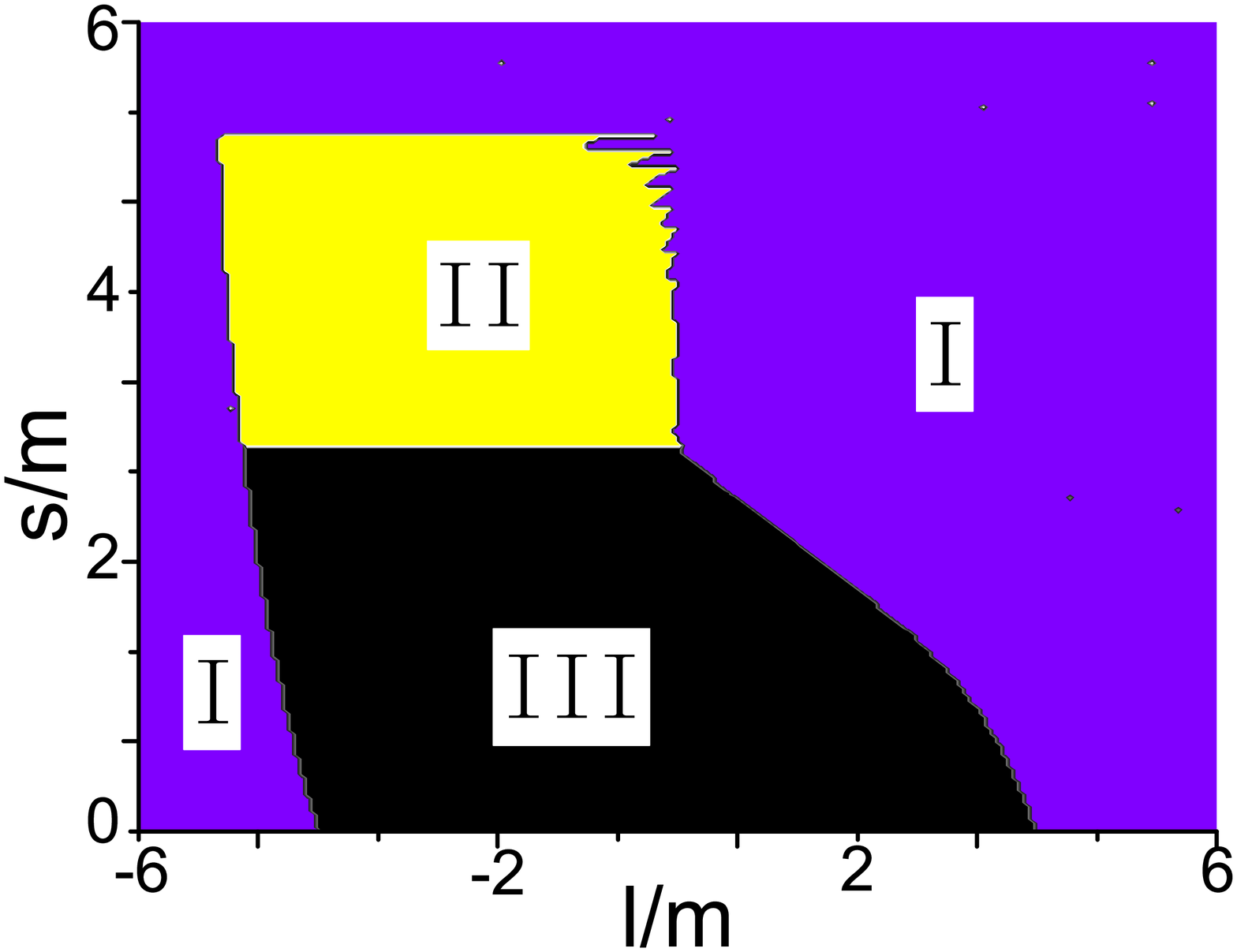}}
    \subfigure[~$Q=0.2$]{
    \includegraphics[ width=0.23\textwidth]{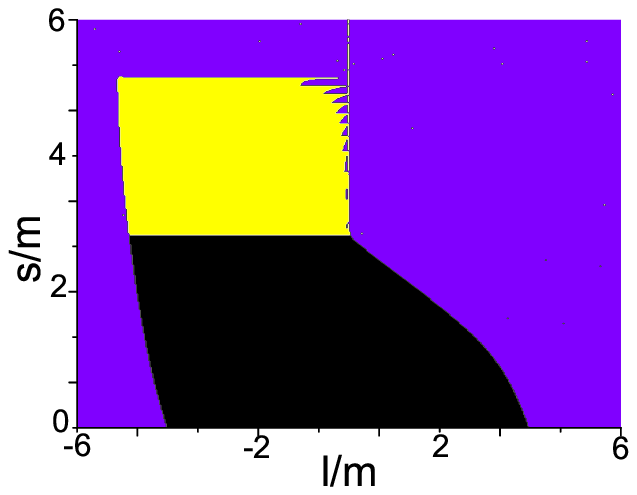}}
    \subfigure[~$Q=0.4$]{
    \includegraphics[ width=0.23\textwidth]{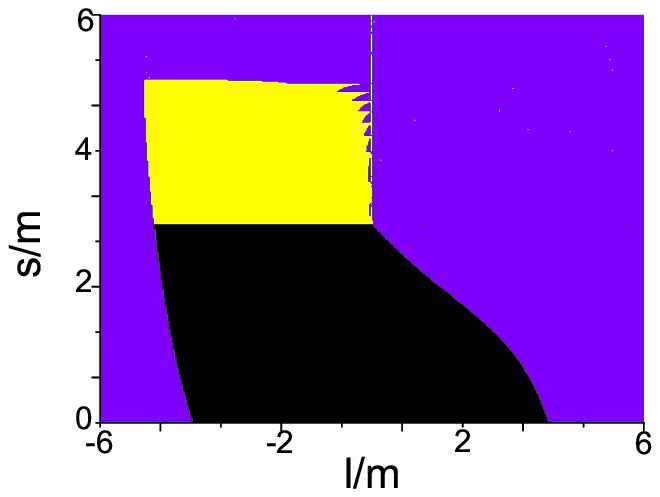}}
    \subfigure[~$Q=0.6$]{
    \includegraphics[ width=0.23\textwidth]{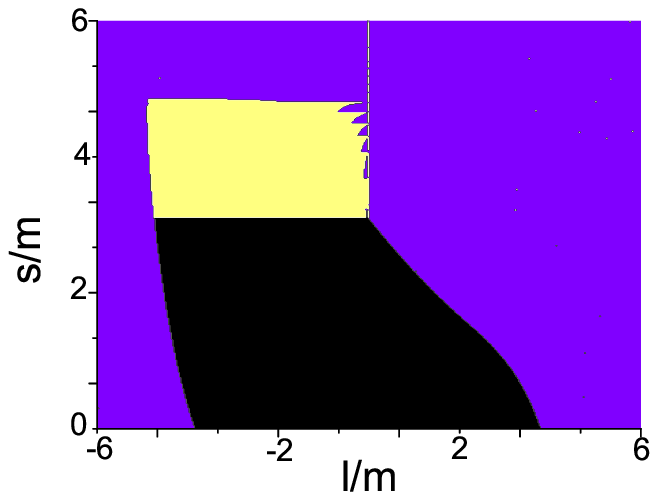}}
    \subfigure[~$Q=0.8$]{
    \includegraphics[ width=0.23\textwidth]{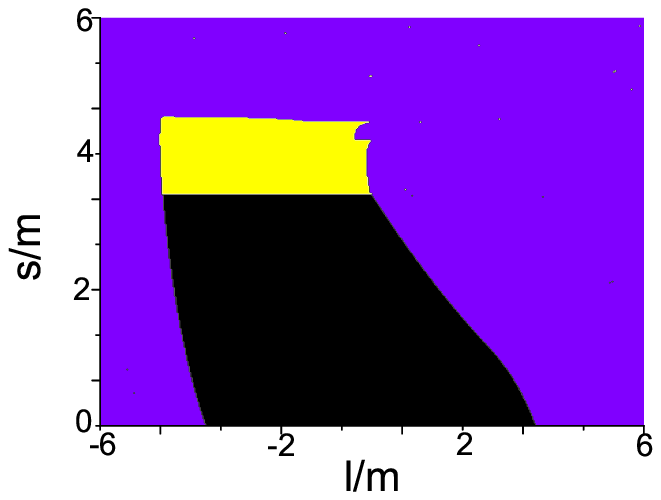}}
    \subfigure[~$Q=1$]{
    \includegraphics[ width=0.23\textwidth]{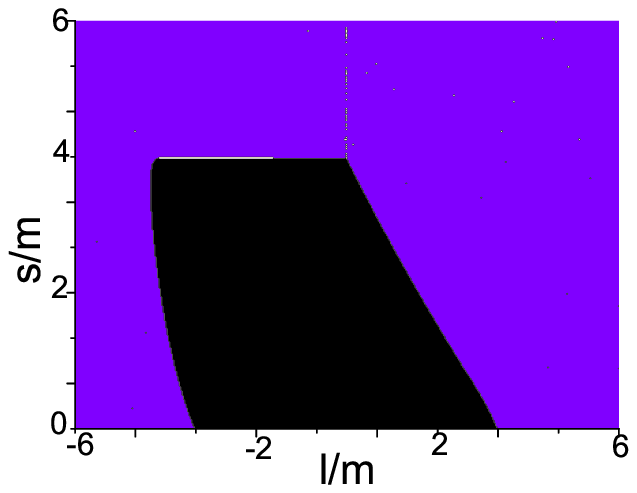}\label{rnturningpointf}}
    \vskip -4mm \caption{Plots of the characters of the CM energy for RN black hole with different charge $Q$ with respect to the orbital angular momentum $l$ and spin the $s$. In region \textrm{I} (purple region) the particles first reach the turning radius,  in region \textrm{II} (yellow region) the divergent radius first, and in region \textrm{III} (black region) the horizon first. The CM energy might be divergent in region \textrm{II}. (The set is the same in the following figures.) The step length of spin and orbital angular momentum are $\frac{6}{400}$ and $\frac{12}{400}$, respectively. The particles move from $r=80$ to the horizon of the black hole and the step length is $\frac{80}{12000}$.}
    \label{nKNQ0SJ}
    \end{figure}

Although the CM energy for the spinning test particles is finite, Eq. (\ref{centerofmassenergyrn})  might still indicate that the CM energy {might} be divergent out of the horizon. Next we investigate the characters of the CM energy out of the horizon for the RN black hole with different values of the total orbital angular momentum $l$ ($l=j-s$) and spin $s$. The corresponding numerical results are shown in Fig. \ref{rnturningpointf}.
We find that the area of the  divergent region in the ($s,~l$) space nearly disappears when the RN black hole charge $Q$ approaches 1. So it is also interesting to study how the charge of the RN black hole affects the CM energy of the spinning test particles. We consider six cases with $Q=0,~0.2,~0.4,~0.6,~0.8,~1$ and give the corresponding numerical results in Fig. \ref{nKNQ0SJ}.
The numerical results in Fig. \ref{nKNQ0SJ} indicate that for the non-extremal RN black hole, the CM energy { might} be divergent if the test particles have spin, and the charge of the RN black hole $Q$ will affect the motion of the particles and the divergent region of the CM energy in the ($s,~l$) space. We can make a brief summary that the change of the RN black hole charge $Q$ can yield the following results:

(1) The area of the  (black) region that particles can reach the horizon in the ($s,~l$) space increases with the charge of the RN black hole $Q$. The corresponding left and right boundaries of the black region move to the center ($l=0$) with the increase of $Q$, which means that the bound of the orbital angular momentum that particles can reach the horizon of the RN black decreases with the increase of the charge.

(2) The area of the  divergent (yellow) region of the CM energy in the ($s,~l$) space decreases with the increase of the charge of the RN black hole. The boundary of the yellow and black regions moves upward with the increase of the charge.

(3) When the black hole charge $Q$ equals $1$ the divergent region vanishes, which is the same as the result in Eq. (\ref{extremalrnhorizon}).

\section{Kerr black hole case}\label{scheme3}

\begin{figure}[!htb]
    \includegraphics[width=0.23\textwidth]{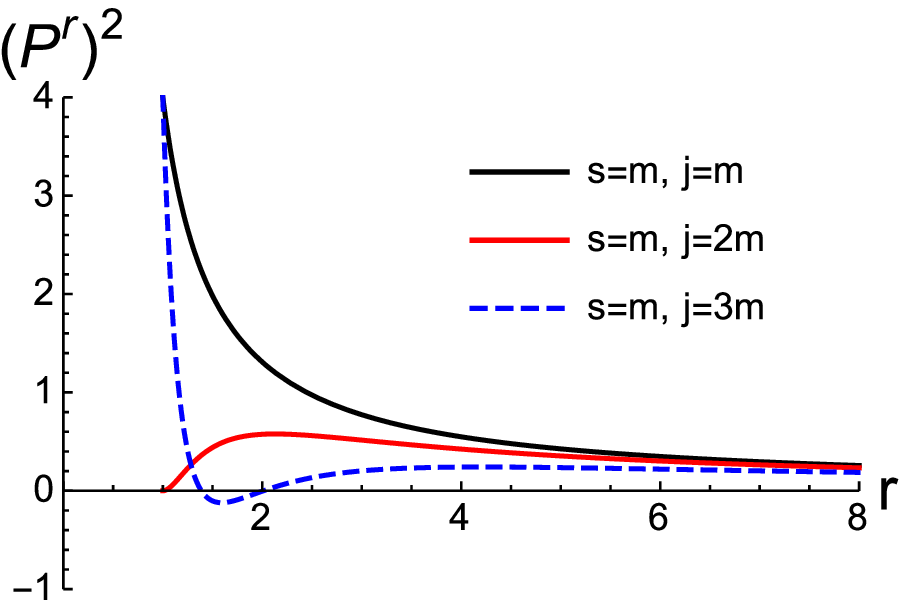}
    \includegraphics[width=0.23\textwidth]{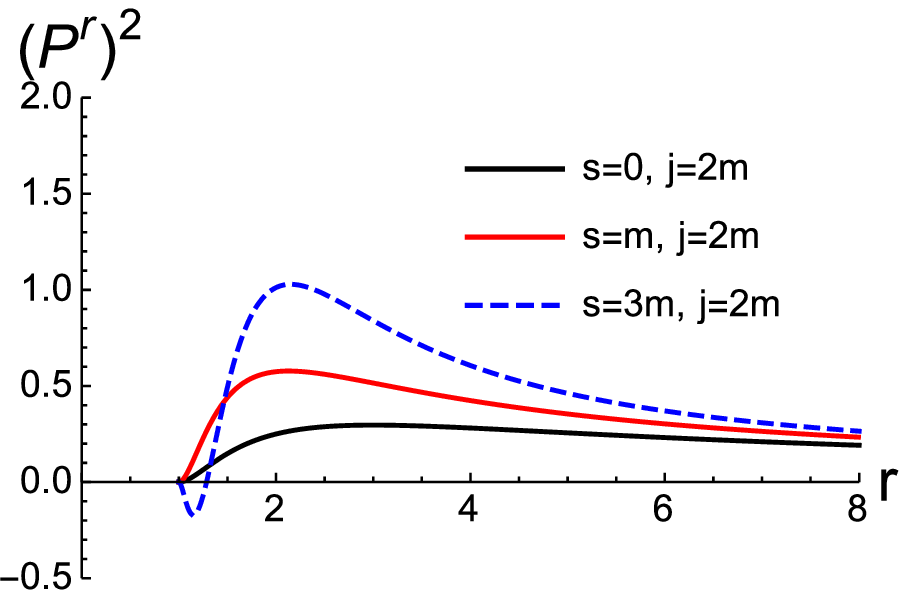}
    \vskip -4mm \caption{Plots of the $(P^r)^2$ with different $s$ and $j$ for an extremal Kerr black hole ($a=1$).}
    \label{turnpointextremalkerr}
    \end{figure}

In this section, we would like to consider Kerr black holes as spinning test particles accelerators, so we set the charge Q=0. Then the CM energy (\ref{centerofmassenergyjiexijie}) reads as
\begin{eqnarray}
(E_{\text{cm}}^{\text{kerr}})^2&=&\frac{2K_{2}}{(m^2 r^4-r m^2\bar{s}_1^2)(m^2 r^4-r m^2\bar{s}_2^2)}\nonumber\\
&&\times\frac{1}{r \left(a^2+(r-2) r\right)^2},\label{centerofmassenergykerr}
\end{eqnarray}
where the $K_2$ is defined in appendix A. The divergent radius $r_d$ based on the spin $\bar{s}$ is
\begin{equation}
r_d=\left(\bar{s}\right)^{\frac{2}{3}},
\end{equation}
For a Kerr black hole the corresponding CM energy near the horizon is
\begin{eqnarray}
(E_{\text{cm}}^{\text{kerr}})^2|_{r\to r_+}&=&\frac{m^2K_{3}}{m^6(1-\bar{s}_1)(1+\bar{s}_1)^2(1-\bar{s}_2)(1+\bar{s}_2)^2}\nonumber\\
&&\times\frac{1}{m^2(2-\bar{j}_1)(2-\bar{j}_2)},\label{cmekerrnearhorizon}
\end{eqnarray}
where the $K_3$ is defined in appendix A. We can see that when the total angular momentum $j_1$ or $j_2$ equals $2m$ the CM energy near the horizon will be divergent. This is consistent with the results in Refs. \cite{Banados2009,Guo2016}. However, in Ref. \cite{Guo2016} the authors did not consider the effects of the spines $s_1$ and $s_2$. We find that when the spines of the particles $s_i=\pm m$ ($i=1,2$), the CM energy near the horizon might be divergent. This is a new feature, and should not be ignored. Now we check exactly that, for the critical values of the spin $s$ and total angular momentum $j$, whether the particles can reach the horizon. Here we also use the zero point of $P^r$ to define the turning radius. For an extremal Kerr black hole the figure of $(P^r)^2$ with $s=m$ and $j=2m$ is shown in Fig. \ref{turnpointextremalkerr}. Clearly we see that the particles can reach the horizon and the CM energy can be divergent near the horizon.

\begin{figure}[!htb]
    \subfigure[$~a=0$]{
    \includegraphics[width=0.23\textwidth]{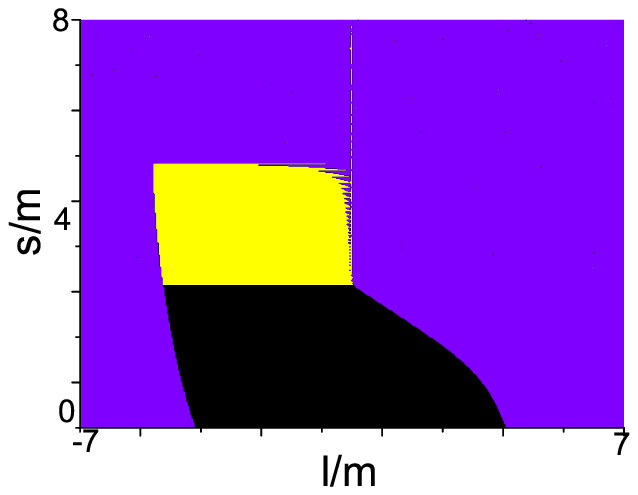}}
    \subfigure[$~a=0.2$]{
    \includegraphics[width=0.23\textwidth]{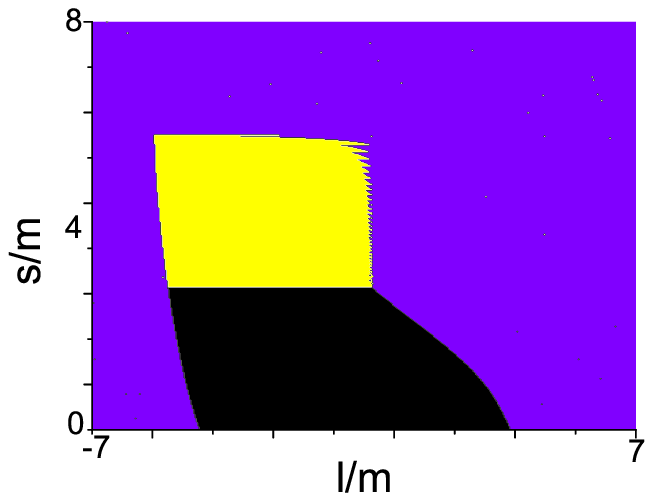}}
    \subfigure[$~a=0.4$]{
    \includegraphics[width=0.23\textwidth]{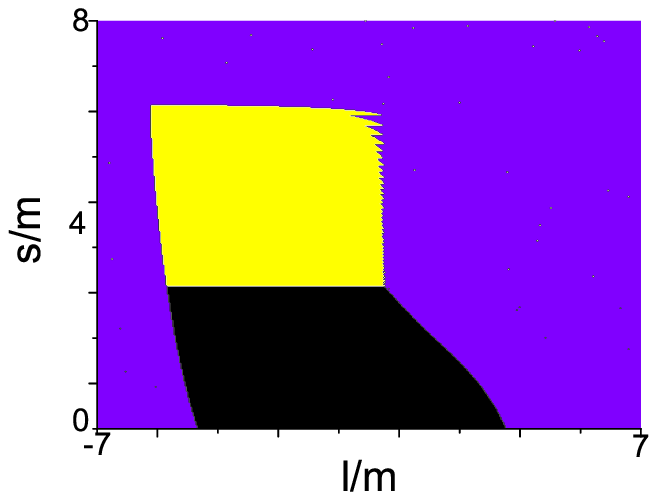}}
    \subfigure[$~a=0.6$]{
    \includegraphics[width=0.23\textwidth]{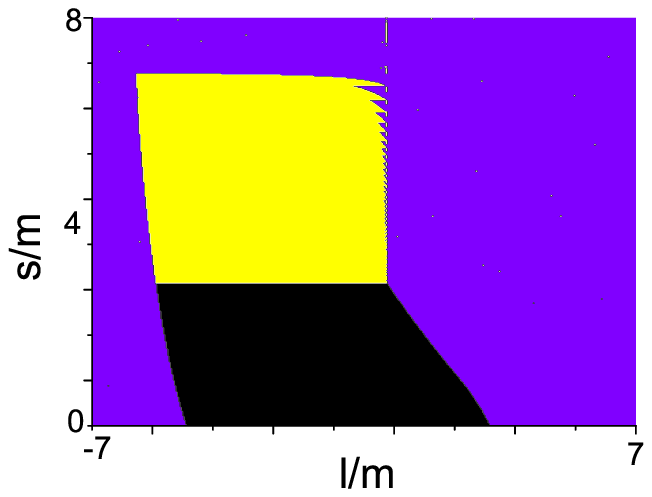}}
    \subfigure[$~a=0.8$]{
    \includegraphics[width=0.23\textwidth]{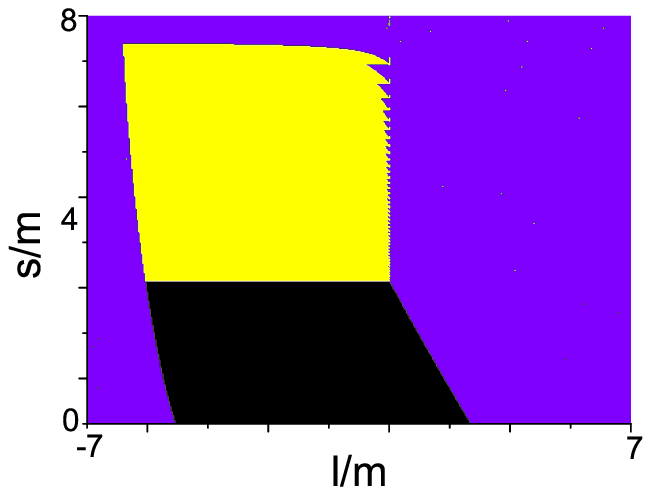}}
    \subfigure[$~a=1$]{
    \includegraphics[width=0.23\textwidth]{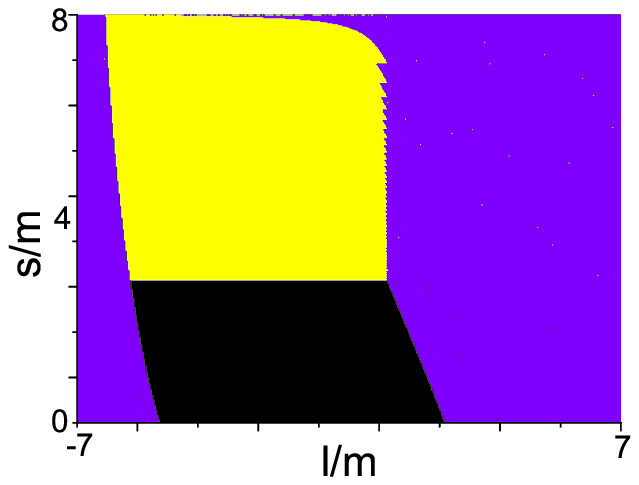}}
    \vskip -4mm \caption{Plots of the characters of the CM energy for a Kerr black hole with different angular momentum $a$ with respect to the orbital angular momentum $l$ and the spin $s$.
    The step length of spin and orbital angular momentum are $\frac{6}{400}$ and $\frac{12}{400}$, respectively. The particles move from $r=80$ to the horizon of the black hole and the step length is $\frac{80}{12000}$.
    }
    \label{kerrnocetremalsj}
    \end{figure}

It is known that astrophysical limitation requires the spin $a<0.0998$ for a Kerr black hole\cite{Thorne1974}. So the more practical case is to consider non-extremal Kerr black holes, and we would like to investigate the spinning test particles colliding in the non-extremal Kerr black hole background. We also use the numerical method to determine the parameter space ($s$ and $l$) in which the CM energy {might} be divergent. The characters of the CM energy with respect to the black hole spin are shown in Fig. \ref{kerrnocetremalsj}. Here the black hole spin is set to be $a=0,~0.2,~0.4,~0.6,~0.8,~1$.

We can make a conclusion that for a Kerr black hole the CM energy of the test spinning particles can be divergent with the astrophysical limitation based on Ref.~\cite{Thorne1974}, and the divergent region of the CM energy in the ($s,~l$) space is affected by the spin of the Kerr black hole. The numerical results in Fig. \ref{kerrnocetremalsj} indicate that the following conclusions:

(1) The area of the divergent (yellow) region of the CM energy in the ($s,~l$) space increases with the spin $a$ of the Kerr black hole.

(2) The boundary of the divergent (yellow) region of the CM energy and the region of particles that can reach the horizon (black region) in the ($s,~l$) space is almost unaffected by the change of the Kerr black hole spin.

(3) In the Kerr black hole background, the range of the orbital angular momentum that a test particle can reach the horizon depends on the direction of its orbital angular momentum in comparison with the direction of the black hole rotation. The co-rotation and counter-rotation cases corresponding to the two directions are parallel or antiparallel. Our numerical result shows that the left and right boundaries of the black region that particles can reach the horizon move a bit to the left with the increase of the Kerr black hole spin $a$ in the ($s,~l$) space. This is consistent with the drag effect of the Kerr black hole.

\section{Extremal Kerr-Newman black hole case}\label{scheme4}

The CM energy for the extremal KN black hole near the horizon is
    \begin{eqnarray}
    (E_{\text{cm}}^{\text{KN}})^2|_{r\to r_+}\!\!\!&=&\!\!\!\frac{K_{4}}{m^2(1\!-\!a\bar{s}_1) (1\!+\!a\bar{s}_1)^2 (1\!-\!a\bar{s}_2) (1\!+\!a\bar{s}_2)^2}\nonumber\\
    &&\times\frac{1}{m^2(a^2-\bar{j}_1 a+1)(a^2-\bar{j}_2 a+1)},\label{cmenergyextremalknhorizon}
    \end{eqnarray}
where the $K_4$ is defined in appendix A. We can see that when $a^2-\bar{j}_1 a+1=0$ or $a^2-\bar{j}_2 a+1=0$ the CM energy will be divergent and the critical total angular momentum $\bar{j}_{i}=\frac{1 + a^2}{a}$ ($i=1,2$), which is the same as that of Ref. \cite{Wei2010}. When { the spines of  the particles satisfy} $\bar{s}_i=\pm\frac{1}{a}$ ($i=1,2$), the CM energy can also be divergent near the horizon. Clearly Eq. (\ref{cmenergyextremalknhorizon}) will reproduce the Eq. (\ref{extremalrnhorizon}) with $a=0$.

\begin{figure}[!htb]
    \includegraphics[width=0.23\textwidth]{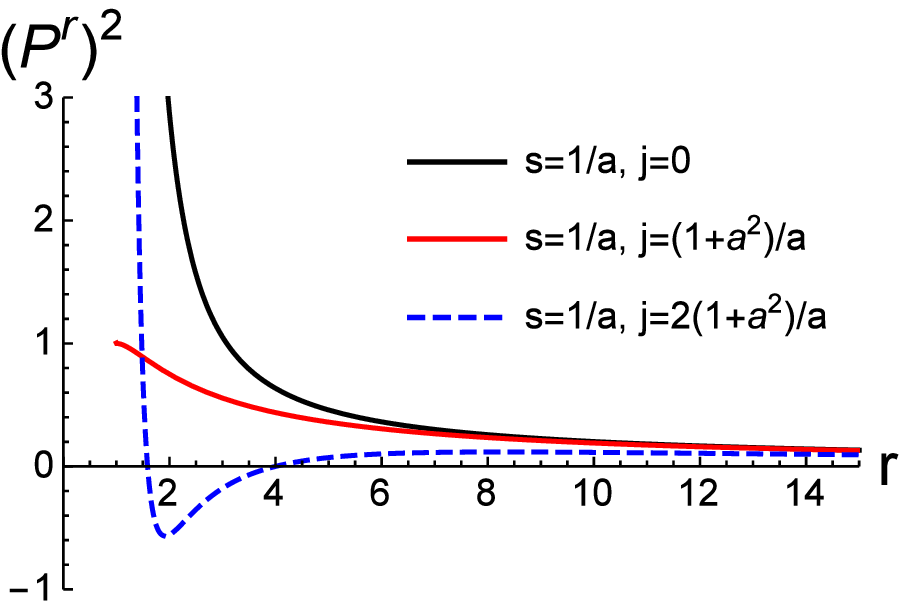}
    \includegraphics[width=0.23\textwidth]{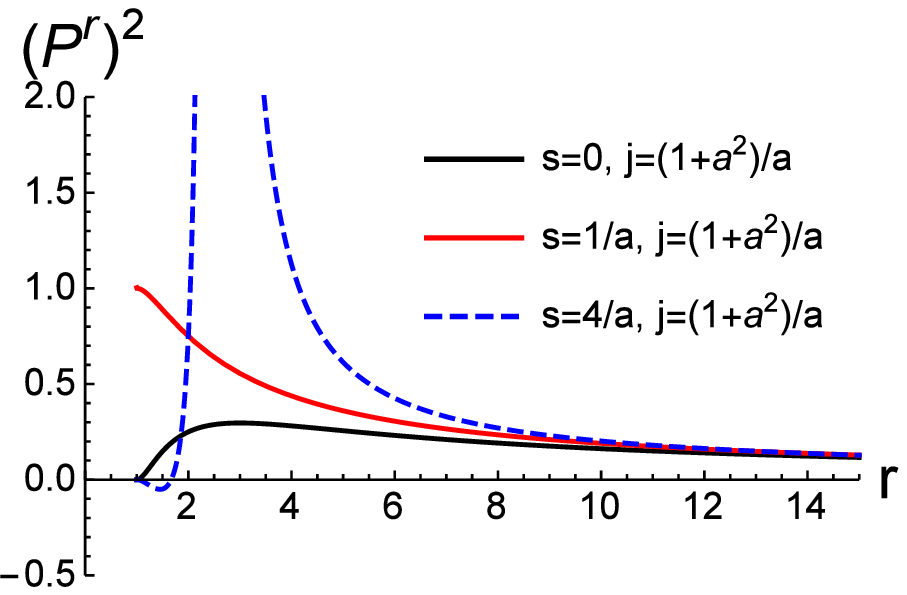}
    \vskip -4mm \caption{Plots of the $(P^r)^2$ with different $s$ and $j$ and for an extremal KN black hole ($a=1$).}
    \label{turnpointextremalkn}
    \end{figure}

\begin{figure}[!htb]
    \subfigure[~$a=0~(Q=1)$ ]{
    \includegraphics[width=0.23\textwidth]{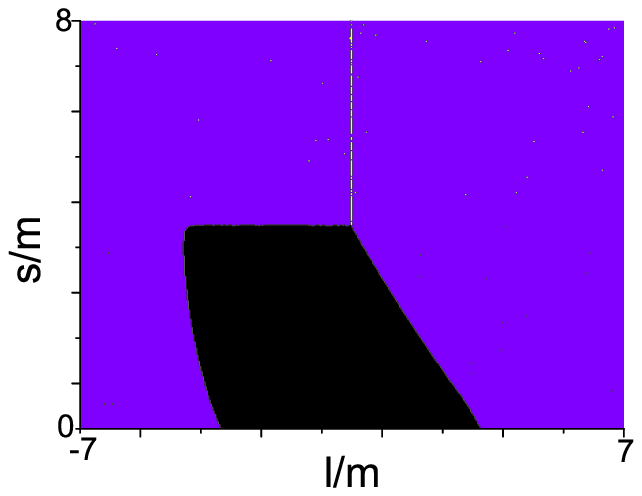}\label{knturningpointA}}
    \subfigure[~$a=0.2~(Q=\sqrt{1-0.2^2})$ ]{
    \includegraphics[width=0.23\textwidth]{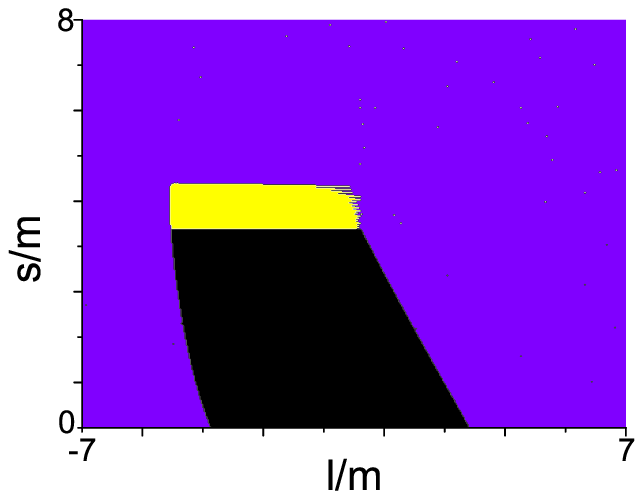}}
    \subfigure[~$a=0.4~(Q=\sqrt{1-0.4^2})$ ]{
    \includegraphics[width=0.23\textwidth]{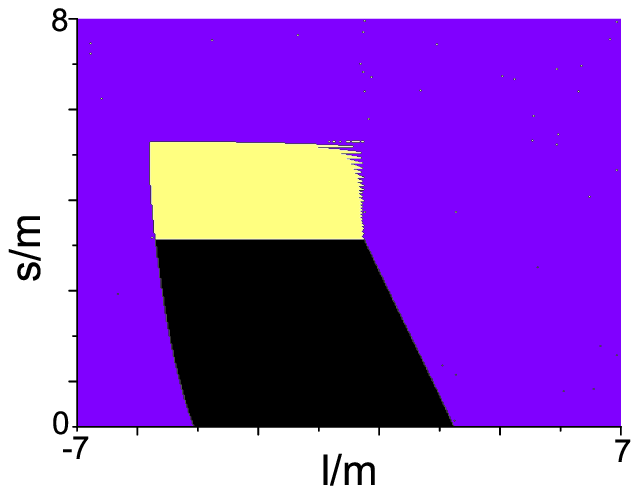}}
    \subfigure[~$a=0.6~(Q=0.8)$ ]{
    \includegraphics[width=0.23\textwidth]{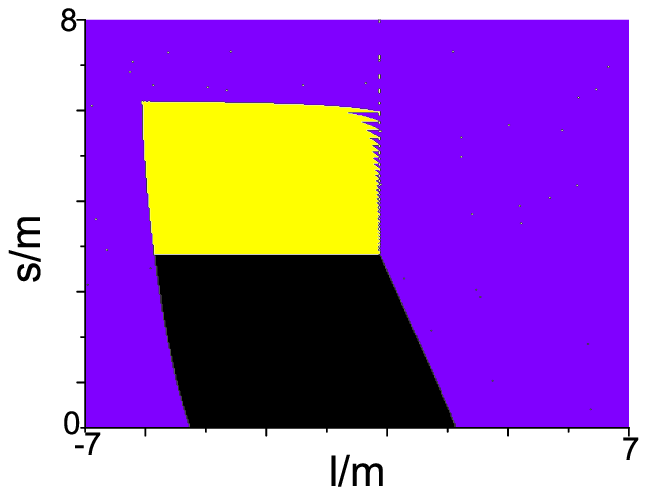}}
    \subfigure[~$a=0.8~(Q=0.6)$ ]{
    \includegraphics[width=0.23\textwidth]{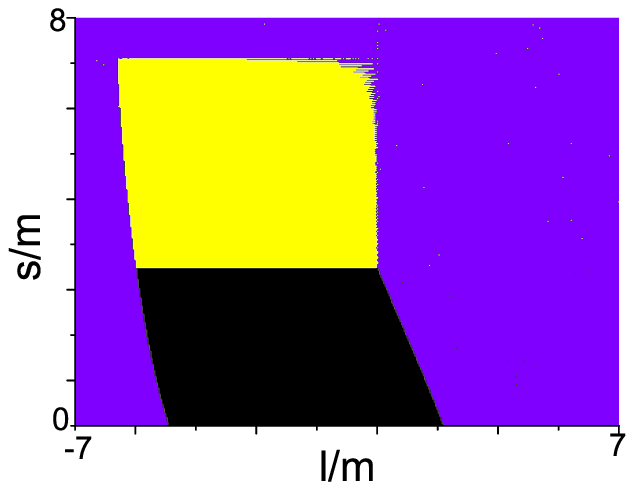}}
    \subfigure[~$a=1~(Q=0)$ ]{
    \includegraphics[width=0.23\textwidth]{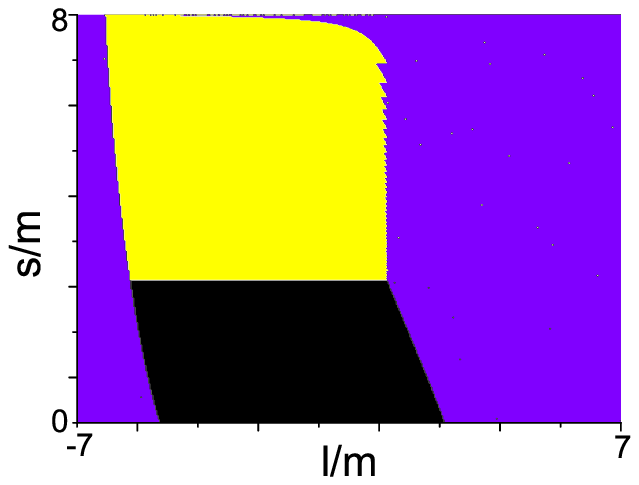}}
    \vskip -4mm \caption{Plots of the characters of the CM energy for an extremal Kerr-Newman black hole with different angular momentum $a$ with respect to the orbital angular momentum $l$ and the spin $s$.
    The step length of spin and orbital angular momentum are $\frac{8}{400}$ and $\frac{14}{400}$, respectively. The particles move from $r=100$ to the horizon of the black hole and the step length is $\frac{100}{15000}$.}
    \label{knturningpoint}
    \end{figure}

Next we should make sure that the particles can reach the horizon when the total angular momentum $\bar{j}=\frac{1+a^2 }{a}$ or spin angular momentum $\bar{s}=\frac{1}{a}$. We use the zero point of $(P^r)^2$ to define the turning point radius. We give the plots of the $(P^r)^2$ with different values of $s$ and $j$ in Fig. \ref{turnpointextremalkn}.
We also use the numerical method to investigate the effects of the black hole spin $a$, where we choose $a=0,~0.2,~0.4,~0.6,~0.8,~1$ for simplicity, and the corresponding numerical results are shown in Fig. \ref{knturningpoint}.

It is easy to see that Fig.~\ref{knturningpointA} is the same as Fig.~\ref{rnturningpointf} since both of them correspond to the case of an extremal KN black hole. When the spin of an extremal KN black hole increases the charge will decrease, so the change of the extremal KN black hole spin should have some properties like the RN and Kerr black holes. The numerical results in Fig. \ref{knturningpoint} indicate the following conclusions:

(1) { The area of the divergent (yellow) region}  of the CM energy in the ($s,~l$) space increases with the spin of an extremal KN black hole, which is consistent with the Kerr black hole case.

(2) The boundary of the divergent (yellow) region of the CM energy and the (black) region of particles that can reach the horizon in the ($s,~l$) space move downward with the increase of the extremal Kerr black hole spin or the decrease of the extremal Kerr black hole charge. This result is similar to the case of the RN black hole.

(3) The left and right boundaries of the black region move a bit to the left with the increase of the extremal black hole spin $a$, which is consistent with the Kerr black hole case.

\section{Velocity of spinning test particles}\label{scheme5}

{ Note that the velocity vector $u^\mu$ and the canonical momentum vector $P^\mu$ of the spinning test particle are not parallel \cite{phdthesis,Armaza2016,Hojman2013,Deriglazov2015,Deriglazov2015a}, and the canonical momentum $P^\mu$ satisfies $P^\mu P_\mu=-m^2$ which indicates that the canonical momentum vector keeps timelike along the trajectory. However, the velocity vector $u^\mu$ of the spinning test particle might transform to be spacelike from timelike \cite{phdthesis,Armaza2016,Hojman2013}. This can be avoided if one considers the reaction of the spinning test particles to the spacetime, and the results will be more accurate. See some more details in Refs. \cite{Deriglazov2015,Deriglazov2015a}. For simplicity, we do not consider the reaction in this paper. We should be aware that in Schwarzschild case in Ref. \cite{Armaza2016} the divergent region of the CM energy in ($s, l$) space was covered by the superluminal region. That is to say the spinning test particles can not reach the divergent radius in Schwarzschild background. Similarly, our extending numerical results about the divergence of CM energy in ($s, l$) space might be unphysical.

Therefore, it is necessary to investigate the relations of the divergent region and superluminal region in ($s, l$) space. Because the velocity vector $u^\mu$ of the spinning test particle might transform to be spacelike from timelike \cite{phdthesis,Armaza2016,Hojman2013,Deriglazov2015,Deriglazov2015a}, we can use the velocity vector $u^\mu$ to determine whether the movement of particles is spacelike or timelike. The corresponding velocity $u^\mu$ can be solved by using the equations of motion (\ref{equationmotion1}) and (\ref{equationmotion2}) based on \cite{Hojman2013} as follows
\begin{eqnarray}
\frac{DS^{tr}}{D\lambda}&=&P^t\dot{r}-P^r\nonumber\\
&=&\frac{S^{\phi r}P_\phi}{P_t^2}\frac{DP_t}{D\lambda}-\frac{DS^{\phi r}}{D\lambda}\frac{P_\phi}{P_t}-\frac{S^{\phi r}}{P_t}\frac{DP_\phi}{D\lambda}\label{velocityequation1}
\end{eqnarray}
and
\begin{eqnarray}
\frac{DS^{t\phi}}{D\lambda}&=&P^t\dot{\phi}-P^\phi\nonumber\\
&=&-\frac{S^{\phi r}P_r}{P_t^2}\frac{DP_t}{D\lambda}+\frac{DS^{\phi r}}{D\lambda}\frac{P_\phi}{P_t}+\frac{S^{\phi r}}{P_t}\frac{DP_r}{D\lambda}\label{velocityequation2}.
\end{eqnarray}
Here the non-zero components of the spin tensor $S^{\mu\nu}$ are
\begin{eqnarray}
S^{r\phi}&=&-S^{\phi r}=-\frac{s P_{t}}{m r},\nonumber\\
S^{rt}&=&-S^{tr}=-S^{r\phi}\frac{P_\phi}{P_t}=s\frac{P_\phi}{mr},\label{spinnozo}\\
S^{\phi t}&=&-S^{t\phi}=S^{r\phi}\frac{P_r}{P_t}=-s\frac{P_r}{mr}.\nonumber
\end{eqnarray}
By inserting Eq. (\ref{spinnozo}) into Eqs. (\ref{velocityequation1}) and (\ref{velocityequation2}) we can obtain
\begin{eqnarray}
\frac{DS^{tr}}{D\lambda}&=&P^t\dot{r}-P^r\nonumber\\
&=&-\frac{s}{mr}\frac{DP_\phi}{D\lambda}+s\frac{P_\phi}{mr^2}\dot{r}\label{spinvelocityequation1}
\end{eqnarray}
and
\begin{eqnarray}
\frac{DS^{t\phi}}{D\lambda}&=&P^t\dot{r}-P^\phi\nonumber\\
&=&\frac{s}{mr}\frac{DP_r}{D\lambda}-s\frac{P_r}{mr^2}\dot{r}\label{spinvelocityequation2}.
\end{eqnarray}
The velocity can be solved by using Eqs. (\ref{spinvelocityequation1}) and (\ref{spinvelocityequation2}) \footnote{We find that the term $\frac{P_\mu}{m~r^2}\frac{Dr}{D\lambda}$ may be lost in Refs. \cite{Armaza2016,Hojman2013} when calculated the derivative term $\frac{D}{D\lambda}\left(\frac{P_\mu}{mr}\right)$. But our result for the superluminal region of a spinning particle for Schwarzschild black hole is the same as Ref.~\cite{Armaza2016}.}.
Finally, we have
\begin{eqnarray}
\frac{u^\mu u_\mu}{(u^t)^2} =
  \frac{g_{tt}}{c^2}
  + g_{rr}\Big(\frac{\dot{r}}{c}\Big)^2
  + g_{\phi\phi}\Big(\frac{\dot{\phi}}{c}\Big)^2  
  + 2g_{\phi t}\dot{\phi}.  ~~~\label{velocitysquare}
\end{eqnarray}

Next we give the numerical result for Schwarzschild black hole in Fig. \ref{velocityspacelikesw}. We can conclude that the CM energy of the spinning test particles can not diverge in the Schwarzschild background by comparing Fig. \ref{nKNQ0SJ} and Fig. \ref{velocityspacelikesw}, which is consistent with the result in Ref. \cite{Armaza2016}. We also give the numerical results for RN, Kerr, and KN black holes in Figs. \ref{rnvelocityspacelike}, \ref{kerrvelocityspacelike}, and \ref{knvelocityspacelike}, respectively.

    \begin{figure}[!htb]
    \includegraphics[width=0.23\textwidth]{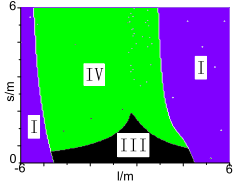}
    \vskip -4mm \caption{Plots of the characters of the trajectories for a spinning test particle in Schwarzschild background as a function of the orbital angular momentum $l$ and the spin $s$. The step length of spin and orbital angular momentum are $\frac{8}{200}$ and $\frac{12}{200}$, respectively. In region \textrm{I} (purple region) the particles first reach the turning radius,  in region \textrm{IV} (green region) the superluminal radius first, and in region \textrm{III} (black region) the horizon first. Here the region \textrm{IV} stands for the motion of the spinning test particle is spacelike (i.e., the superluminal region) (The set is the same in the following figures). The particle moves from $r=80$ to the horizon of the black hole and the step length is $\frac{80}{40000}$.}
    \label{velocityspacelikesw}
    \end{figure}

    \begin{figure}[!htb]
    \subfigure[~$Q=0.5$ ]{
    \includegraphics[width=0.23\textwidth]{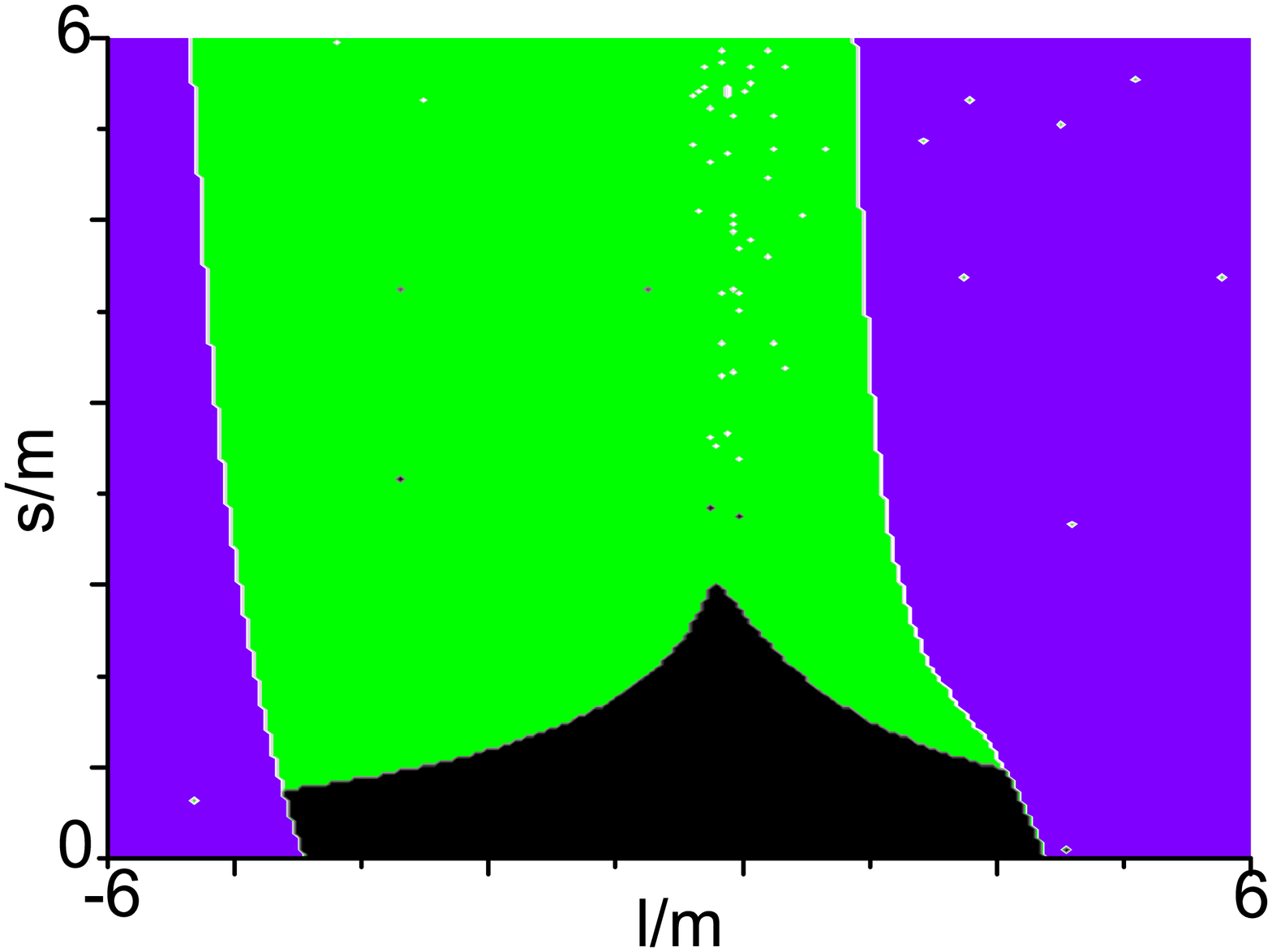}\label{rnspacelikea}}
    \subfigure[~$Q=1$ ]{
    \includegraphics[width=0.23\textwidth]{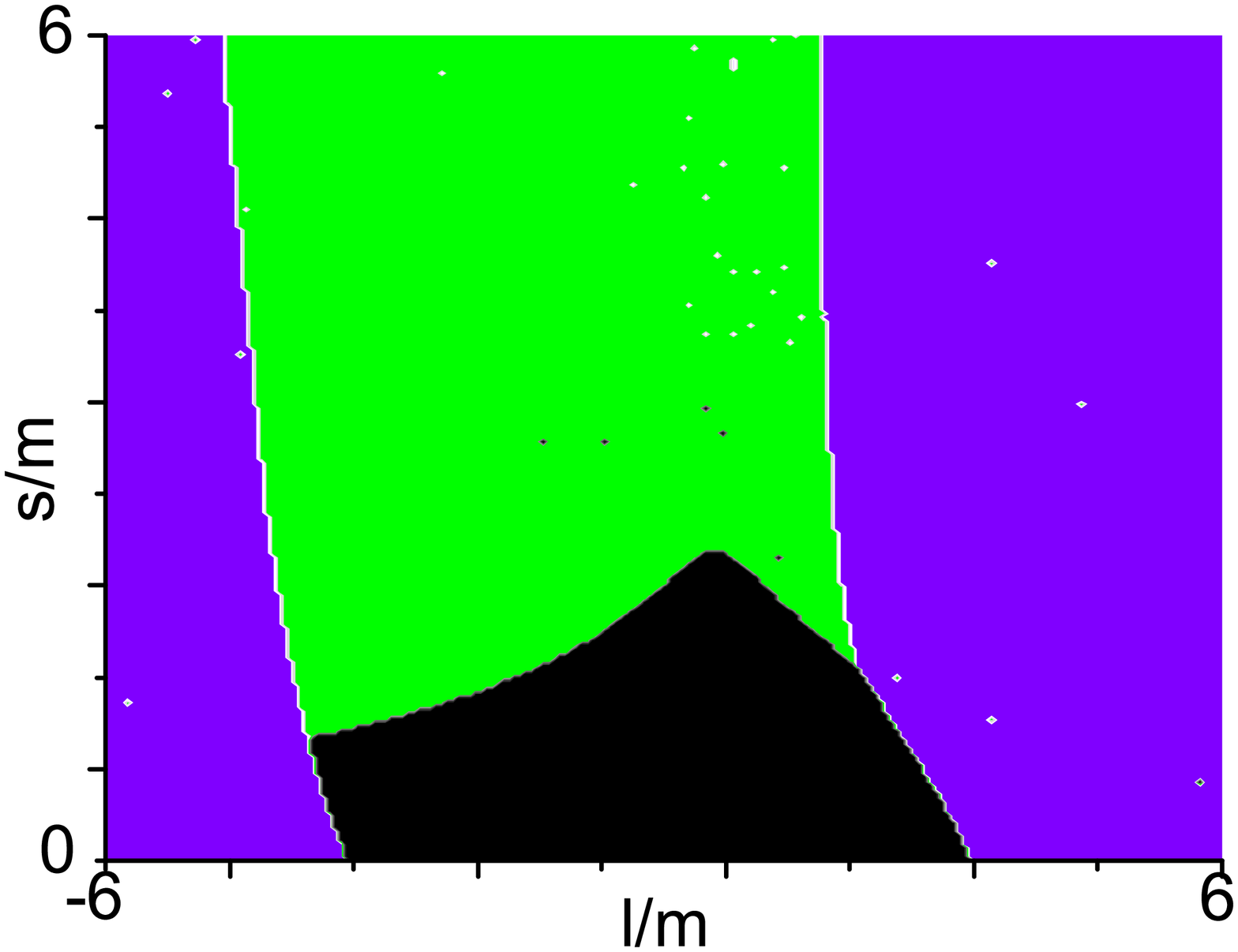}\label{rnspacelikeb}}
    \vskip -4mm \caption{Plots of the characters of the trajectories for a spinning test particle in RN background with different values of the charge $q$ as a function of the orbital angular momentum $l$ and the spin $s$. The step length of spin and orbital angular momentum are $\frac{6}{200}$ and $\frac{12}{200}$, respectively. The particle moves from $r=80$ to the horizon of the black hole and the step length is $\frac{80}{24000}$.}
    \label{rnvelocityspacelike}
    \end{figure}
    \begin{figure}[!htb]
    \subfigure[~$a=0.25$ ]{
    \includegraphics[width=0.23\textwidth]{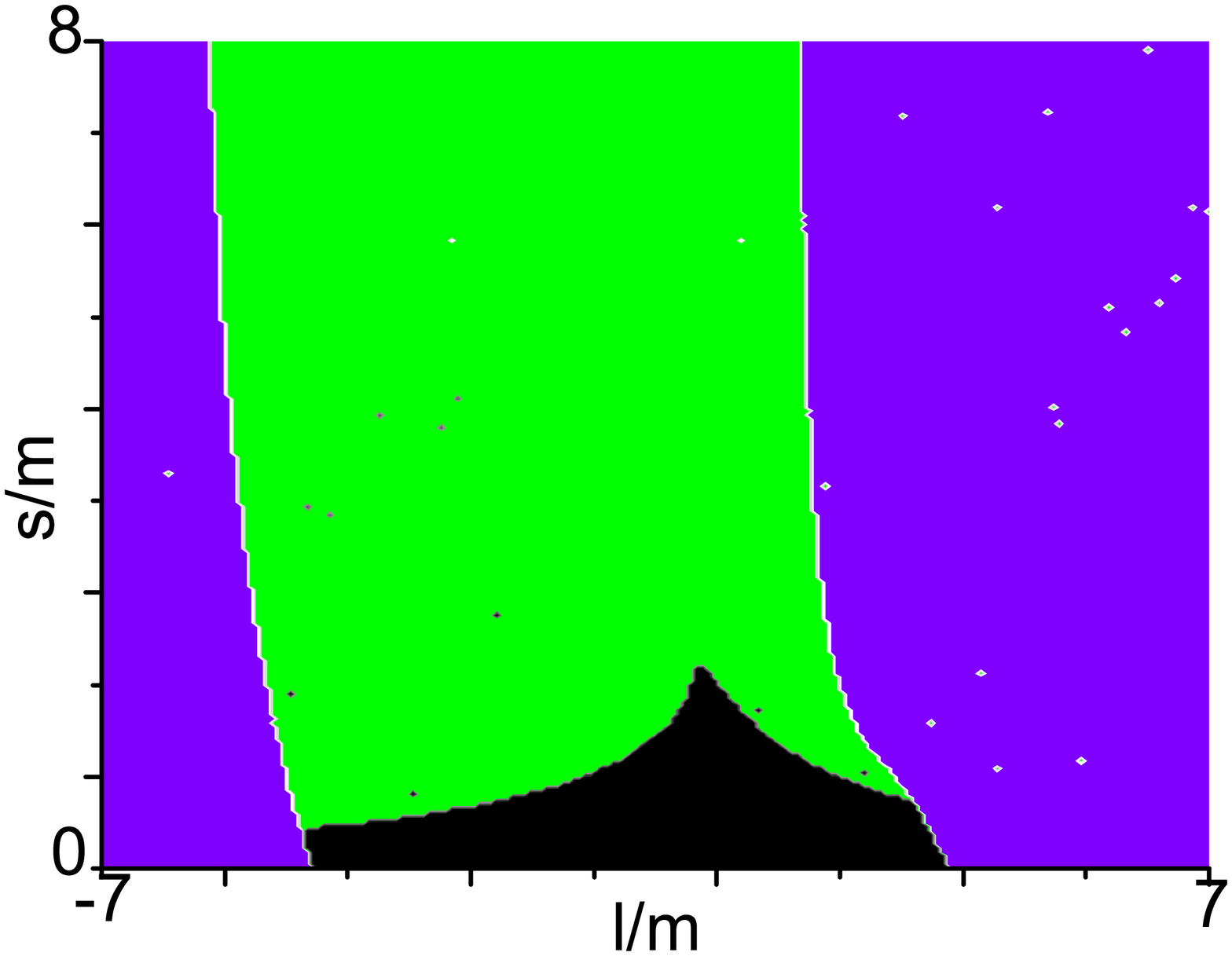}\label{rnspacelikea}}
    \subfigure[~$a=0.75$ ]{
    \includegraphics[width=0.23\textwidth]{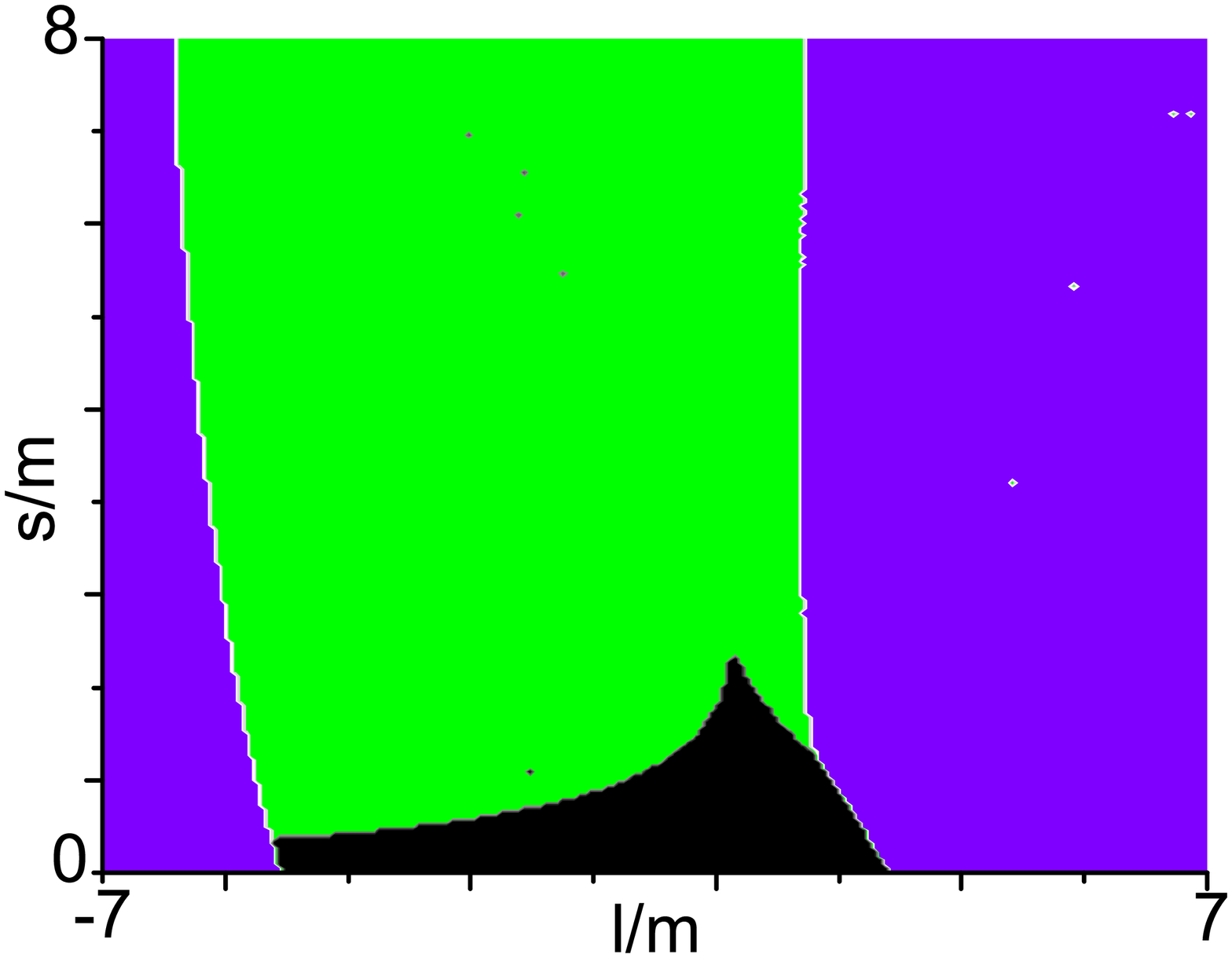}\label{rnspacelikeb}}
    \vskip -4mm \caption{Plots of the characters of the trajectories for a spinning test particle in Kerr background with different values of the angular momentum $a$ as a function of the orbital angular momentum $l$ and the spin $s$. The step length of spin and orbital angular momentum are $\frac{8}{200}$ and $\frac{14}{200}$, respectively. The particle moves from $r=100$ to the horizon of the black hole and the step length is $\frac{100}{20000}$.}
    \label{kerrvelocityspacelike}
    \end{figure}
    \begin{figure}[!htb]
    \subfigure[~$a=0.25,~Q=\sqrt{1-0.25^2}$ ]{
    \includegraphics[width=0.23\textwidth]{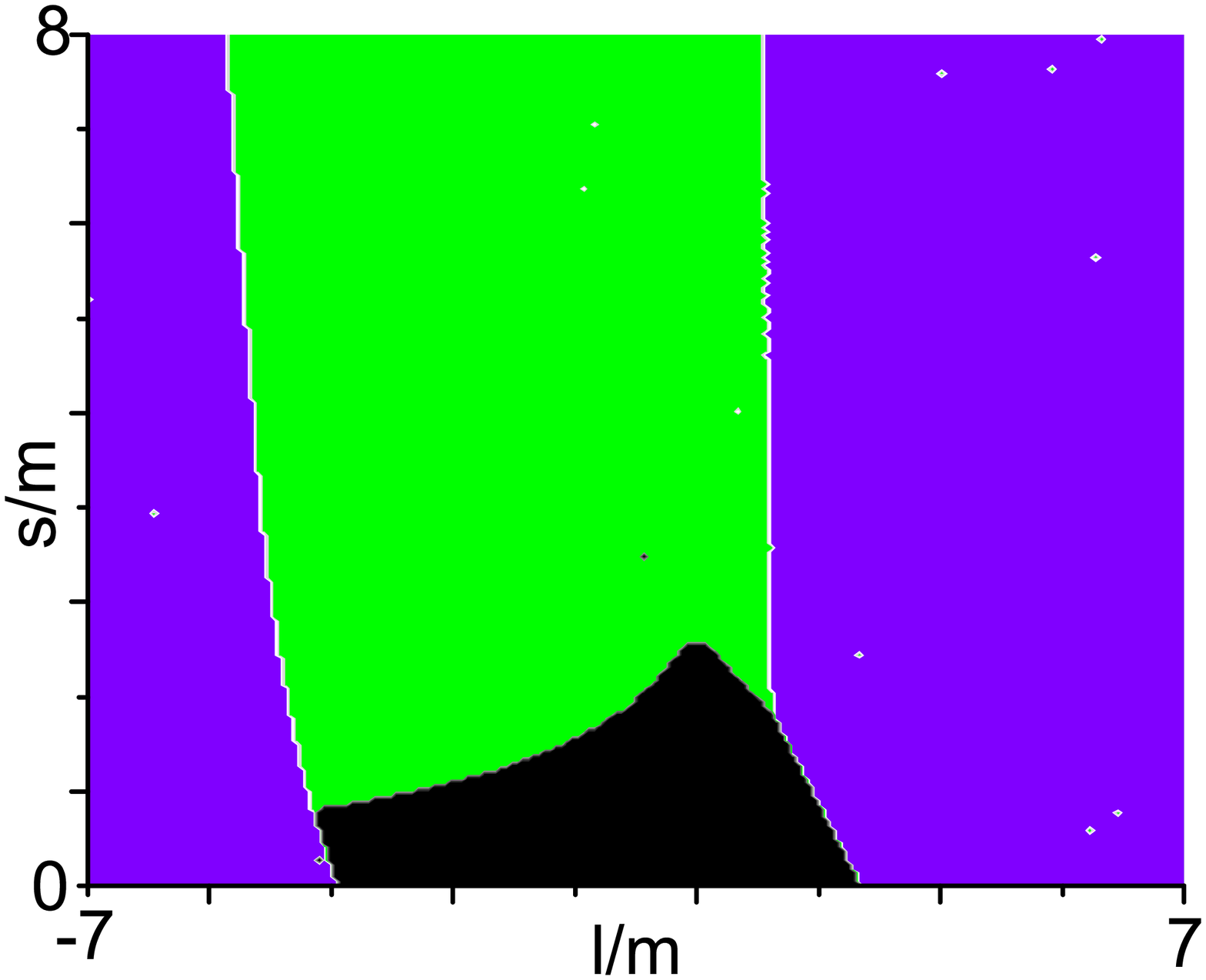}\label{rnspacelikea}}
    \subfigure[~$a=0.5,~Q=\sqrt{1-0.5^2}$ ]{
    \includegraphics[width=0.23\textwidth]{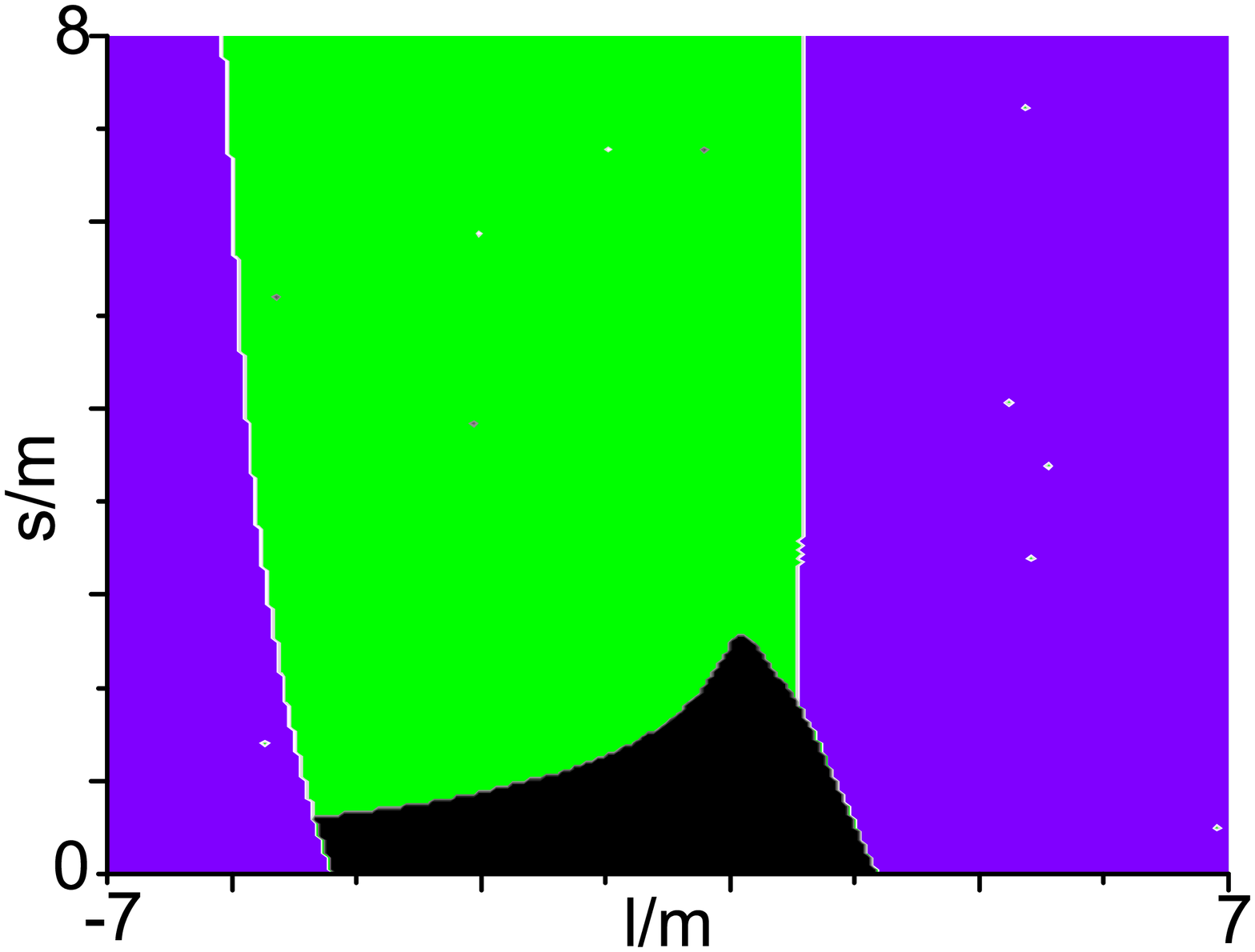}\label{rnspacelikeb}}
    \subfigure[~$a=0.75,~Q=\sqrt{1-0.75^2}$ ]{
    \includegraphics[width=0.23\textwidth]{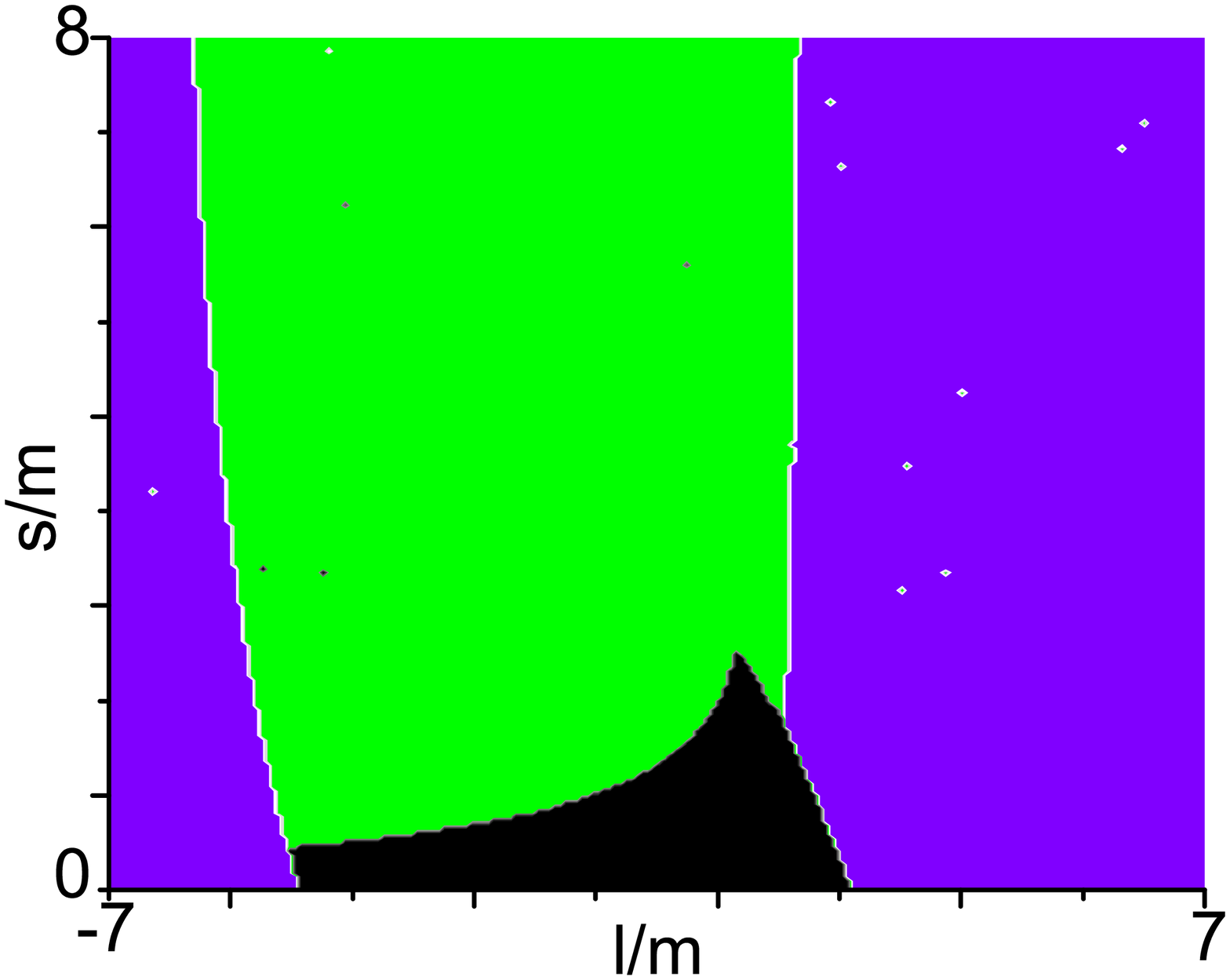}\label{rnspacelikea}}
    \subfigure[~$a=1,~Q=0$ ]{
    \includegraphics[width=0.23\textwidth]{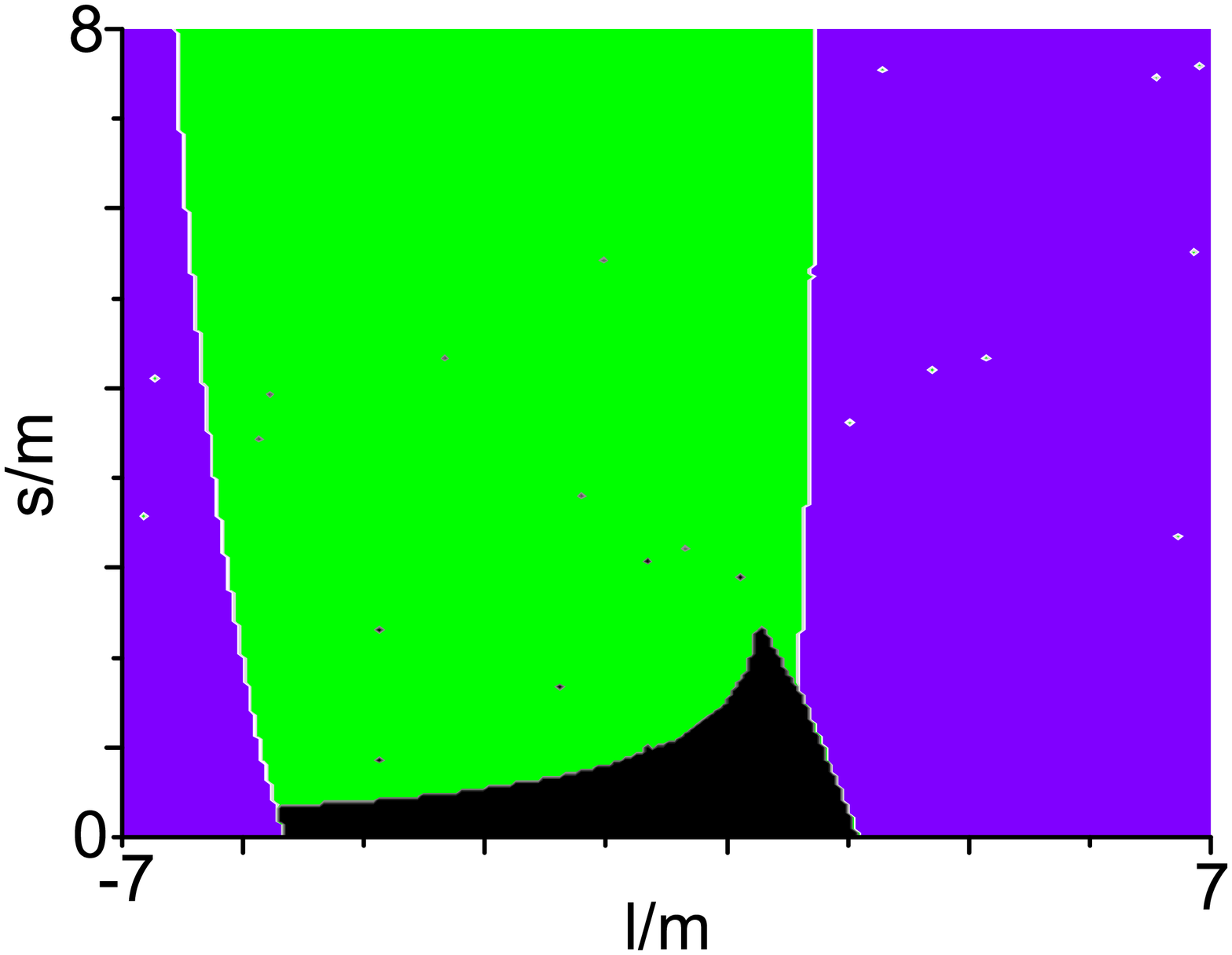}\label{rnspacelikeb}}
    \vskip -4mm \caption{Plots of the characters of the trajectories for a spinning test particle in extremal Kerr-Newman background with different values of the angular momentum $a$ as a function of the orbital angular momentum $l$ and the spin $s$. The step length of spin and orbital angular momentum are $\frac{8}{200}$ and $\frac{14}{200}$, respectively. The particle moves from $r=100$ to the horizon of the black hole and the step length is $\frac{100}{20000}$.}
    \label{knvelocityspacelike}
    \end{figure}

It is easy to see that the divergent region of the CM energy all be covered by the superluminal region in ($s, l$) space by comparing the divergent region and superluminal region, which means that the spinning test particles actually can not reach the divergent radius out of the horizon.}
\section{Summary and conclusion}\label{Conclusion}

In this paper we first reviewed the Lagrangian to solve the equations of motion for classical spinning test particles in curved spacetime and obtained the corresponding four momenta and the CM energy of two spinning test particles in a general KN black hole background. We investigated the possibility that RN ($a=0$), Kerr ($Q=0$), and extremal KN black holes as spinning test particle accelerators by the equations of motion for classical spinning test particles.

We showed that for the extremal RN black hole case the CM energy near the horizon is finite for two spinning particles with arbitrary spin and total angular momentum, which is similar to the Schwarzschild black hole case. For a non-extremal RN black hole, if the spins of the particles are not zero, the CM energy {might} be divergent out of the event horizon and the divergence of the CM energy is affected by the charge of the black hole. The area of the divergent region in the $(s,l)$ space decreases with the black hole charge. For the extremal Kerr black hole case, it was found that the CM energy of two spinning particles can be divergent when the total angular momentum $j$ of one of the two particles reaches its critical value $j_c=2m$, which is the same as the case without particle spin. Furthermore, if the spin of one of the particles satisfied $s=m$, the CM energy {might} be divergent. For a non-extremal Kerr black hole, the CM energy {might} be divergent out of the horizon and the divergent region in the $(s,l)$ space increases with the spin of black hole. For an extremal KN black hole, we found that when the total angular momentum $j$ of one of the particles satisfied $j=j_c=\frac{m+a^2m}{a}$ or the spin of the particle satisfied $s=\frac{m}{a}$, the CM energy {might} be divergent near or out of the horizon of the extremal KN black hole. The area of the  divergent region in the $(s,l)$ space increases/decreases with the spin/charge of the extremal KN black hole.

It was found that the influences of the spin and charge of these black holes are completely opposite for the CM energy of two spinning test particles. More interestingly, compared with the spinless cases, for which the CM energy can only be divergent at the horizon, the CM energy of two spinning particles might be divergent outside the horizon, which indicates that the spinning test particles might approach arbitrary high energy within finite time. { But we should note that the velocity vector $u^\mu$ of the spinning test particle might transform to be spacelike from timelike when it moves from infinity to the horizon of black hole, so Eqs. (\ref{equationmotion1}) and (\ref{equationmotion2}) to describe the motion of the spinning test particles are only applicable at low energy limit. By comparing the divergent region and the superluminal region in ($s,~l$) space, we finally confirmed that the divergent radius of the CM energy out of the horizon can not be reached because of the motion is spacelike.}

\acknowledgments{
We thank Professor Sergio A. Hojman for sharing his PhD thesis to help us to solve the equations of motion of the spinning test particles in curved spacetime. { We would also like to thank the referee for pointing out that the relationship between the divergent region and the superluminal region in ($s,~l$) space should be clarified}. This work was supported in part by the National Natural Science Foundation of China (Grants Nos. 11205074, 11375075, and 11522541)}, and the Fundamental Research Funds for the Central Universities (Grants Nos. lzujbky-2015-jl1, lzujbky-2016-k04, and lzujbky-2016-121).

\section*{Appendix A} 

This appendix contains the complete definitions of $K_1$, $K_2$, and $K_3$ appeared in Eqs. (\ref{centerofmassenergyrn}), (\ref{centerofmassenergykerr}), and (\ref{cmekerrnearhorizon}). The expressions of $K$ and $K_4$ in Eqs.~(\ref{centerofmassenergyjiexijie}) and (\ref{cmenergyextremalknhorizon}) are too long, so we do not list them.
\begin{widetext}
\begin{eqnarray}
K_1&=&{m^2 r^2 \left(m^2\bar{j}_1 \bar{s}_1 \left(Q^2-r\right)+m^2 r^4\right) \left(m^2\bar{j}_2 \bar{s}_2 \left(Q^2-r\right)+m^2 r^4\right)}-m^6 r^6 \left(\bar{j}_1-\bar{s}_1\right) (\bar{j}_2-\bar{s}_2)\left(Q^2+(r-2) r\right)\nonumber\\
    &&-m \bigg[2 \bar{j}_1 m^4 r^6 \bar{s}_1 \left(2 Q^2+(r-3) r\right)+m^2\bar{j}_1^2 r^2 \left(m^2\bar{s}_1^2 (Q^2-r)^2-m^2 r^4 \left(Q^2+(r-2) r\right)\right)\nonumber\\
    &&+m^4 r^8 (2 r-Q^2)-m^2\bar{s}_1^2 \left(Q^2+(r-2) r\right) \left(m^2 r^4 \left(2 Q^2+(r-2) r\right)+m^2\bar{s}_1^2 (Q^2-r)^2\right)\bigg]^{1/2}\nonumber\\
    &&\times m\bigg[ 2 \bar{j}_2 m^4 r^6 \bar{s}_2 \left(2 Q^2+(r-3) r\right)+m^2\bar{j}_2^2 r^2 \left(m^2\bar{s}_2^2 (Q^2-r)^2-m^2 r^4 \left(Q^2+(r-2) r\right)\right)\nonumber\\
    &&+m^4 r^8 (2 r-Q^2)-m^2\bar{s}_2^2 \left(Q^2+(r-2) r\right) \left(m^2 r^4 \left(2 Q^2+(r-2) r\right)+m^2\bar{s}_2^2 (Q^2-r)^2\right)\bigg]^{1/2}\nonumber\\
    &&+\left(Q^2+(r-2) r\right) \left(m^2 r^4+m^2\bar{s}_1^2 (Q^2-r)\right) \left(m^2 r^4+m^2\bar{s}_2^2 (Q^2-r)\right)m^2,
\end{eqnarray}
\begin{eqnarray}
K_2&=&r \left(a^2+(r-2) r\right)^2 (m^2 r^4-r m^2\bar{s}_1^2)(m^2 r^4-r m^2\bar{s}_2^2) m^2-\left(a^2+(r-2) r\right) h_1~h_2\nonumber\\
&&+(r-2) r^2 \left(m \left(m \left(r^3+a^2 (r+2)\right) r^2+\left(a^3+3 r^2 a\right) m\bar{s}_1\right)-m\bar{j}_1 \left(2 a m r^2+(a^2+r^2) m\bar{s}_1\right)\right)\nonumber\\
 &&\times\left(m \left(m \left(r^3+a^2 (r+2)\right) r^2+(a^3+3 r^2 a) m\bar{s}_2\right)-m\bar{j}_2 \left(2 a m r^2+(a^2+r^2) m\bar{s}_2\right)\right) m^2\nonumber\\
 &&-r^2 \left(r^3+a^2 (r+2)\right) \left(m \left(2 a M r^2+\left(a^2-(r-2) r^2\right) m\bar{s}_1\right)+m\bar{j}_1 \left(m (r-2) r^2-a m\bar{s}_1\right)\right)\nonumber\\
  &&\times\left(m \left(2 a m r^2+\left(a^2-(r-2) r^2\right) m\bar{s}_2\right)+m\bar{j}_2 \left(m (r-2) r^2-a m\bar{s}_2\right)\right) m^2\nonumber\\
  &&+2a\bigg[\left[m\bar{j}_1 \left(2 a m r^3+(a^2+r^2) m\bar{s}_1 r\right)-m r \left(m \left(r^3+a^2 (r+2)\right) r^2+(a^3+3 r^2 a) s_1\right)\right] \nonumber\\
  &&\times\left[-2 a m^2 r^3+(r-2) m^2(\bar{s}_2-\bar{j}_2) r^3-a^2 m^2 \bar{s}_2 r+a m^2\bar{j}_2 \bar{s}_2 r\right]\nonumber\\
  &&+\left[-2 a m^2 r^3+(r-2) m^2(\bar{s}_1-\bar{j}_1) r^3-a^2 m^2 \bar{s}_1 r+am^2\bar{j}_1 \bar{s}_1 r\right] \nonumber\\
  &&\times\left[m\bar{j}_2 \left(2 a m r^3+(a^2+r^2) m\bar{s}_2 r\right)-m r \left(m \left(r^3+a^2 (r+2)\right) r^2+(a^3+3 r^2 a)m\bar{s}_2\right)\right]\bigg],
\end{eqnarray}
\begin{eqnarray}
K_3&=&m^2\bar{j}_1^2 \bigg[\bar{j}_2^2 m^6 (\bar{s}_1-\bar{s}_2)^2+2 \bar{j}_2 m^6 (\bar{s}_1-\bar{s}_2) (1-\bar{s}_1)(1+\bar{s}_2)\nonumber\\
&&+m^2(1+\bar{s}_2)^2 \big(2 m^4 (1+\bar{s}_1^2)-2 m^4 \bar{s}_2 (1+\bar{s}_1)^2+m^4\bar{s}_2^2 (1+\bar{s}_1)^2\big)\bigg]\nonumber\\
&&+m^2(1+\bar{s}_1)^2 \bigg[m^2\bar{j}_2^2 \left(2 m^4 (1+\bar{s}_2^2)-2 m^4\bar{s}_1 (1+\bar{s}_2)^2+m^4\bar{s}_1^2(1+\bar{s}_2)^2\right)\nonumber\\
&&-4 \bar{j}_2 m^2 (m-m\bar{s}_1) (2 m-m\bar{s}_1-m\bar{s}_2) (m+m\bar{s}_2)^2+4 m^2 (m+m\bar{s}_2)^2 (-2 m+m\bar{s}_1+m\bar{s}_2)^2\bigg]\nonumber\\
&&-2 m\bar{j}_1(m+m\bar{s}_1)(m-m\bar{s}_2) \bigg[m^2 \left(4 m^3-m\bar{s}_2 \left(m\bar{j}_2(m+m\bar{s}_2)+m^2\bar{j}_2^2-6 m^2+2 m^2\bar{s}_2^2\right)\right)\nonumber\\
&&+m^2 \bar{s}_1 \left(-\bar{j}_2 m^2(m+m\bar{s}_2)+m^2\bar{j}_2^2 m+2(m-m\bar{s}_2)(m+m\bar{s}_2)^2\right)-s_1^2(2 m-j_2)(m+s_2)^2\bigg],
\end{eqnarray}
where
\begin{eqnarray}h_i&=&\Bigg[M^2 r^2 \bigg(m^2 \left(m^2(\bar{e}-1)(\bar{e}+1) r^3+2 m^2 r^2-\left((m^2-m^2\bar{e}^2) a^2+m^2\bar{j}_i^2\right) r+2 m^2(\bar{j}_i-a \bar{e})^2\right) r^5\nonumber\\
&&+\left(\left(2 r m^2\bar{e}^2+m^2\bar{e}^2+2 m^2 r\right) a^2-2 m^2\bar{e} (r+1) \bar{j}_i a+j_2^2+(r-2) r^2 (2 m^2-e^2 r)\right) s_i^2 r^2\nonumber\\
&&+2 m \left(e j_i r^4+3 m (a e-j_i) r^3+a (j_i-a e)^2 r\right) s_2 r^2-\left(a^2+(r-2) r\right) m^4\bar{s}_i^4\bigg)\Bigg]^{1/2}.\nonumber
\end{eqnarray}
\end{widetext}

\end{document}